\documentclass[useAMS,usenatbib]{mn2e}
\usepackage{graphicx}
\usepackage{textcomp}
\usepackage{txfonts}
\usepackage{comment}
\usepackage{capt-of}
\usepackage[hyperindex,breaklinks=true,colorlinks,citecolor=blue,linkcolor=blue]{hyperref}

\usepackage{subfigure}% in preamble

\usepackage[dvipsnames]{xcolor} %to change colour of corrected parts of the paper
\author[Demetroullas et al.]{C. Demetroullas$^{1}$\thanks{\href{mailto:constantinos.demetroullas@postgrad.manchester.ac.uk}{\nolinkurl{costantinos.demetroullas@postgrad.manchester.ac.uk} }}, C. Dickinson$^{1}$\thanks{\href{mailto:Clive.Dickinson@manchester.ac.uk}{\nolinkurl{Clive.Dickinson@manchester.ac.uk} }}, D. Stamadianos$^{1,3}$, S.E. Harper$^{1}$, \newauthor K. Cleary$^{2}$, Michael E. Jones$^{4}$,  T.J.Pearson$^{2} $, A.C.S.Readhead$^{2}$, Angela C. Taylor$^{4}$.
\\
$^{1}$Jodrell Bank Centre for Astrophysics, Alan Turing building, School of Physics and Astronomy, The University of Manchester, Oxford Road, Manchester, 
\\
M13 9PL, U.K.
\\
$^{2}$Cahill Centre for Astronomy and Astrophysics, California Institute of Technology, Pasadena, CA 91125, USA.
\\
$^{3}$School of Physics, Astronomy and Mathematics, University of Hertfordshire, Hatfield, AL 10 9AB, U.K.
\\
$^{4}$Sub-department of Astrophysics, University of Oxford, Denys Wilkinson Building, Keble Road, Oxford OX1 3RH, U.K.\vspace{-2mm}
}

\date{Accepted xxxx xxxxxxxx xx. Received xxxx xxxxxxxx xx; in original form xxxx xxxxxxxx xx}\vspace{-0mm}

\title{Observations of Galactic star-forming regions with the Cosmic Background Imager at 31\,GHz}

\begin{document}
\pdfpageheight 11.692in

\pagerange{\pageref{firstpage}--\pageref{lastpage}} \pubyear{2012}

\maketitle

\label{firstpage}

\begin{abstract}
Studies of the diffuse Galactic radio emission are interesting both for better understanding the physical conditions in our Galaxy and for minimising the contamination in cosmological measurements. Motivated by this we present Cosmic Background Imager 31\,GHz observations of the Galactic regions NGC\,6357, NGC\,6334, W51 and W40 at $\sim$4$\farcm$5 resolution and conduct an investigation of the spectral emission process in the regions at 4$\farcm$5 and 1$^{\circ}$ resolution. We find that most of the emission in the regions is due to optically thin free-free. For 2 sub-regions of NGC\,6334 and for a sub-region of W51 though, at 4$\farcm$5 resolution and at 31\,GHz we detect less emission than expected from extrapolation of radio data at lower frequencies assuming a spectral index of $-$0.12 for optically thin free-free emission, at 3.3$\sigma$, 3.7$\sigma$ and 6.5$\sigma$ respectively. We also detect excess emission in a sub-region of NCG~6334 at 6.4$\sigma$, after ruling out any possible contribution from Ultra Compact HII (UCHII) regions. At 1$^{\circ}$ resolution we detect a spinning dust component in the Spectral Energy Distribution (SED) of W40 that accounts for 18\,$\pm$\,7\,\% of the total flux density in the region at the peak frequency of 37\,GHz. Comparison with 100\,${\rm \mu m}$ data indicate an average dust emissivity for the sub-regions of $0.5\pm4.4$\,$\mu$K(MJy sr$^{-1}$)$^{-1}$. Finally we translate the excess emission in the regions to an Anomalous Microwave Emission (AME) emissivity relative to the optical depth at 250 ${\rm \mu m }$. We find that this form of emissivity is independent of the AME significance and has a value somewhere in the order of 10$^4$\,Jy.

\end{abstract}

\begin{keywords}
 radiation mechanisms: thermal  - ISM: clouds - HII regions - supernova remnants - radio continuum\vspace{-8mm}
\end{keywords}

\section{Introduction}

The Galactic diffuse emission at radio frequencies provides us with a wealth of information about the formation, structure and evolution of our Galaxy \citep{planck2011b,planck2011}.  
Studies of the Milky Way's radio emission among others enabled us to learn about the nurseries and graveyards of stars \citep{walsh1998,green2009,becker2010,peters2010} and the magnetic field of the Galaxy \citep{mathewson1968,spoelstra1972,vinyaikin1995}. Deciphering the diffuse emission mechanisms of the Galaxy at radio frequencies, also enables us to better probe the CMB fluctuations \citep{jaffe2004,eriksen2006,eriksen2008,bennett2013,planck2013}, thus providing more accurate cosmological information. Knowledge of the spatial morphology and frequency dependence is crucial in an accurate characterization of the foreground emission. Firmly understood components of the diffuse foreground emission include synchrotron, free-free and thermal (vibrational) dust emission. It has been established that free-free is the dominant emission mechanism near the Galactic plane at frequencies between $\sim$\,10 and 100\,GHz \citep{planck2014b}.

In recent years an additional Galactic component has been identified. Anomalous Microwave Emission (AME) has been detected in numerous experiments over the frequency range of $\sim$\,10-60\,GHz, peaking at $\sim$\,30\,GHz in flux density \citep{kogut1996,leitch1997,deOliveiraCosta1997,Lagache2003,casassus2004,Finkbeiner2004,Davies2006,dickinson2007,dickinson2009,todorovic2010,Gold2011,bonaldi2012,tibbs2013}. This new component also shows a strong correlation with far infrared emission, thus inferring a relation to dust grains \citep{leitch1997,finkbeiner2004b,planck2011}.

A number of possible emission mechanisms were proposed over the years to interpret this excess emission in the radio. These include hot ($T_{\rm e}$ $>$~$10{^6}$~K) free-free \citep{leitch1997}, flat synchrotron \citep{bennett2003}, electric dipole radiation from small grains rotating very rapidly (hereafter referred to as spinning dust emission, \citealt{draine1998}) and magnetic dipole radiation from thermal fluctuations in the magnetization of the interstellar grains \citep{draine1999}. However current observations favour the spinning dust model \citep{Watson2005,casassus2008,Scaife2010,planck2011}.

AME has been observed in a number of different environmental conditions throughout the Galaxy. It has been observed in bright HII regions \citep{dickinson2007,dickinson2009}, molecular clouds \citep{Watson2005,casassus2008,Tibbs2010}, planetary nebulae \citep{casassus2004,casassus2007}, dark clouds \citep{Finkbeiner2002,casassus2006,scaife2008,dickinson2010,harper2015}, supernova remnants \citep{Scaife2007}, and photodissociation regions \citep{casassus2008,tibbs2012b}. On the other hand only upper limits have been placed for AME in compact HII regions \citep{scaife2008}.

This paper presents a search for excess emission in a number of bright HII regions and Supernova Remnants (SNRs) near the Galactic plane, at 31\,GHz and $\sim$\,4$\farcm$5 resolution, using the Cosmic Background Imager (CBI). Most of the regions that this paper focusses on were not surveyed in the past to look for excess emission at these frequencies, while the one of them that has (W40) showed tentative evidence of spinning dust \citep{Finkbeiner2004}. Additionally the radio emission in all regions of interest shows a strong correlation with the emission in the far infrared (FIR). By using data from the literature convolved to the CBI resolution we estimate the free-free contribution for the sources at 31\,GHz and by comparing the results with data taken by the CBI we measure or place upper limits on excess emission in some regions while detect an emission deficit in others. We then use ancillary data to form the SED of the regions at 1$^{\circ}$ resolution. Section\,\ref{data} describes the CBI and ancillary data that were used, while Section\,\ref{rr} presents the results for each region. Finally conclusions and discussion are given in Section\,\ref{concl}.

\section{DATA}\label{data}
\subsection{Cosmic Background Imager}
The Cosmic Background Imager (CBI) was an interferometer array, located in Atacama desert in northern Chile at an altitude of 5080\,m \citep{padin2002,taylor2011}. The CBI operated from 1999 to 2008 (CBI1 from 1999 to 2006 and CBI2 from 2006 to 2008). The CBI1 consisted of 13, 0.9\,m antennas that were mounted on a 6\,m tracking platform. The platform could rotate about the optical axis providing improved u-v coverage. The antennas could take measurements either of right (R) or left (L) circular polarisation, allowing observations in total intensity (RR or LL) or polarisation (RL or LR). Each antenna had a receiver operating between 26 and 36\,GHz. This frequency bandwidth was divided into 10 channels of 1\,GHz \citep{padin2002}. The platform supported baselines between 1 and 5.5~m thus providing a maximum resolution of $\sim$\,6$'$, and a primary beam full width at half maximum (FWHM) of 45.2$'$ at the central frequency of 31\,GHz. The nominal system temperature of the telescope was 30\,K. In 2006 the CBI1 was upgraded to CBI2 by replacing the 0.9~m antennas with 1.4~m antennas. This upgrade increased the effective collecting area and allowed observations at higher resolution ($\sim$\,4$\farcm$5) without having to compromise surface brightness sensitivity. The measured CBI2 RMS thermal noise in each frequency channel was at 3.9\,Jy\,s$^{1/2}$\citep{pearson2005,taylor2011}.

\subsection{Observations}

This paper presents the results for the observations at 31\,GHz of the two regions designated as Area-I and Area-II, and the cloud complex W40 (see Table\,\ref{observationsum}). These short time observations were designed to fit between the longer CMB and SZ experiments conducted by the CBI (refs in preparation). 

The two (Area) regions were chosen as they have very bright radio emission that is aligned with the FIR emission, and have not been surveyed in the past at frequencies where AME dominates the spectrum ($\sim$\,30\,GHz). The two regions combined cover an area of $\sim$\,30 deg$^2$. Area-I, consists of 70 individual pointings. The area spans between $349^{\circ}$ and $353^{\circ}50'$ in Galactic longitude ($l$) and $0^{\circ}$ up to $+3^{\circ}$ in Galactic latitude ($b$). Area-I was observed for 10 days between September and October 2007, with a total integration time of $\sim$\,13.5 hours. Area-II consists of 40 individual pointings. It extends between $47^{\circ}50'$ and $51^{\circ}50'$ in $l$ and $-02^{\circ}50'$ up to $00^{\circ}50'$ in $b$. Area-II data were collected in a total of 31 days spread across February and April 2008. The total integration time was $\sim$\,26.5 hours. 
 The mapping strategy for both regions was to observe a number of adjacent fields (10 for Area-I and 8 for Area-II) $0.5^{\circ}$ in diameter, over identical ranges in azimuth and elevation, for short time periods ($\sim$\,2 minutes per field). Each set of adjacent pointings had one or two associated primary flux calibration observations. This strategy was selected to remove the ground spillover that was a problem for the telescope on the shortest baselines.

 The W40 complex has been chosen as previous studies of the region showed indications of anomalous emission; \citet{finkbeiner2004b} reported an emission excess at $\sim$\,33\,GHz as shown in the spectrum of W40 in fig.\,8 of \citet{finkbeiner2004b} where it is evident that the spinning dust model fits the data quite well. W40 is a relatively compact source ($\sim$\,8$'$) centred at (\textit{l},{\textit{b})\,=\,($28\fdg$8,\,$3\fdg$5). The W40 data comprise a total of seven hours of lead/trail scans of the region, bracketed by observations of the primary calibrators. The data were collected on six days between June 2006 and May 2008 \citep{stamadianos2010}.
 
\begin{table*}
\begin{center}

\begin{tabular}{l c c c p{8.0cm}}
    \hline
    \hline
    Region & \textit{l} & \textit{b} & Total Integration time (hr)  & Notes \\
    \hline
    Area-I & $349^{\circ}$ -- $354^{\circ}$ & $0^{\circ}$ -- $3\degr$ & 13.5 & Contains the star forming regions NGC\,6334 and NGC\,6357. \\
    \hline
    Area-II & $48^{\circ}$ -- $ 52^{\circ}$ & $-3\fdg$ -- $1^{\circ}$ & 26.5  & Contains the radio source W51, which in turn is composed of two star-forming regions and an SNR.  \\
    \hline
    W40 &$28\fdg8$ & $3\fdg5$ & 6.7 & Single complex located near the Galactic plane. At radio frequencies W40 is $\sim$\,8\,arcmin across. \\
    \hline
    \end{tabular}
\end{center}

\caption{Summary of the CBI observations.}
\label{observationsum}
\end{table*}

\subsection{Data reduction and Calibration}\label{dataredcal}

The reduction and calibration of the data was performed using the software package CBICAL (developed by T.J. Pearson; see \citealt{readhead2004b,readhead2004} and references therein). The editing and flagging included removing receivers that were malfunctioning, baselines and channels that were contaminated by non-astronomical signal, and data for which the amplitude and phase of the calibrating sources was not constant in time. Absolute flux calibrations were performed using Jupiter (mainly), Saturn or Tau\,A tied to the temperature of Jupiter at 33\,GHz. We assume a temperature $T_{\rm j}^{33GHz}=146.6 \pm 0.8$\,K and a spectral index $\alpha$=$-0.1$ \citep{taylor2011}. The accuracy of the calibrations was limited by telescope phase and amplitude errors and u-v coverage. Comparison of various calibrators over time to estimate the accuracy of the relative flux calibrations indicated that the flux measurements for Area-I, Area-II and W40 were accurate to $\sim$\,6.5\,\%, $\sim$\,10\,\% and $\sim$\,5\,\% respectively.

The dominant source of systematic contamination for the CBI at the level of $\sim$\,0.5\,Jy, especially for the shortest baselines was likely due to radiation from the ground ('ground spillover', \citealt{dickinson2007}). For the Area-I and W40 data-sets the ground spillover was removed by performing lead/trail field observations in a differenced mode. Performing the same technique on the Area-II data-set led to a reduction in the signal-to-noise ratio and dynamic range of the lead fields. Therefore we decided to use the data without performing the ground spillover removal. This is not a significant factor as we have found that for sources at the level of $\gtrsim$10\,Jy (applies for all Area-II sources) the ground signal is negligible.

\subsection{Imaging the Data}\label{dataim}

The reduced and calibrated CBI-2 data were imaged using the software DIFMAP \citep{shepherd1994}. A uniform weighting was selected to provide optimal resolution maps since it was found that our data, due to deconvolution residuals and calibration/pointing errors, were limited by dynamic range at a level of $\sim$\,100:1 (the thermal noise in all maps was $\sim$\,1000:1). Finally the images were primary beam corrected up to the telescope's FWHM. The CBI CLEANed maps of Area-I, Area-II and W40 are shown in Fig.\,\ref{allareas}. The peak 31\,GHz flux for the two bright sources of emission in the Area-I map are 39.2\,Jy/beam and 30.2\,Jy/beam. The maximum flux density in the Area-II map is 64.9\,Jy/beam, while the peak flux value for W40 is at 9.2\,Jy/beam.  The synthesised beam for the Area-I and Area-II and W40 CBI maps at 31\,GHz are $5\farcm4\times 4\farcm2$, $5\farcm9\times 4\farcm8$ and $4\farcm3\times 3\farcm8$, respectively. The measured noise away from any bright sources was found to be $\sim$\,0.05\,Jy/beam, for all three data-sets.

\begin{figure*}
%\hill
\subfigure{\includegraphics[width=5.76cm]{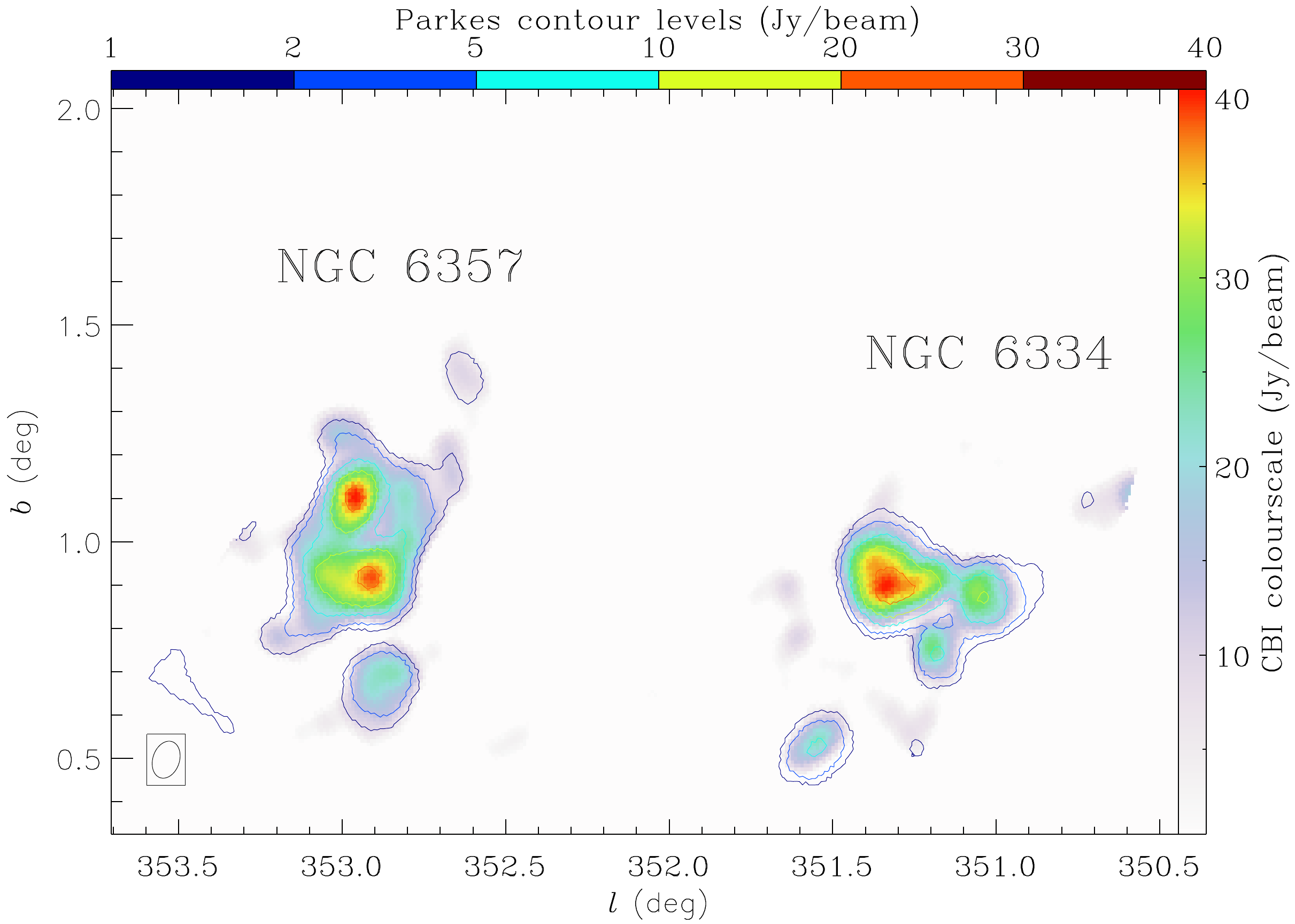}}\hspace{0.5em}
%\hill
\subfigure{\includegraphics[width=5.76cm]{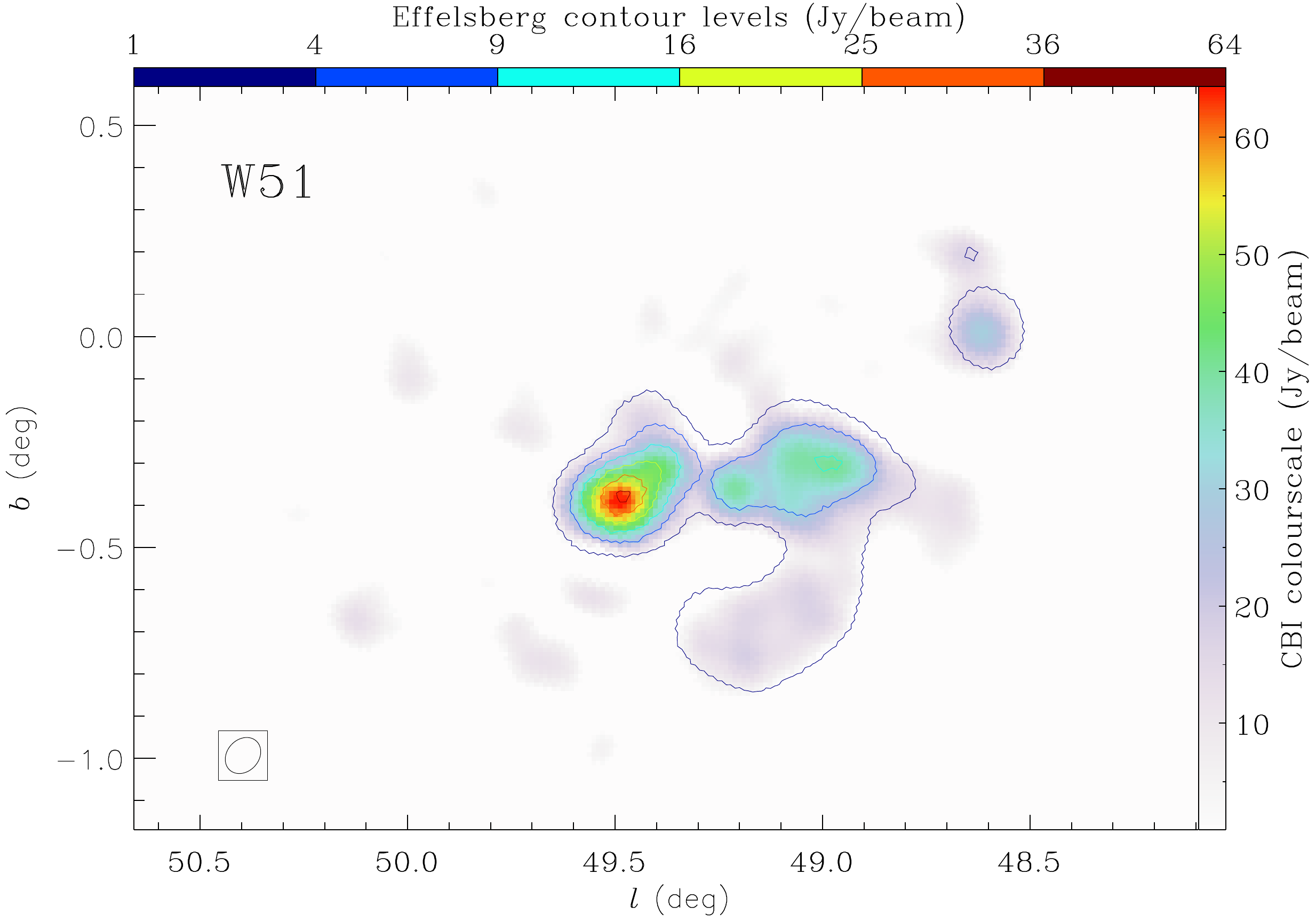}}\hspace{0.5em}
%\hfill
\subfigure{\includegraphics[width=5.76cm]{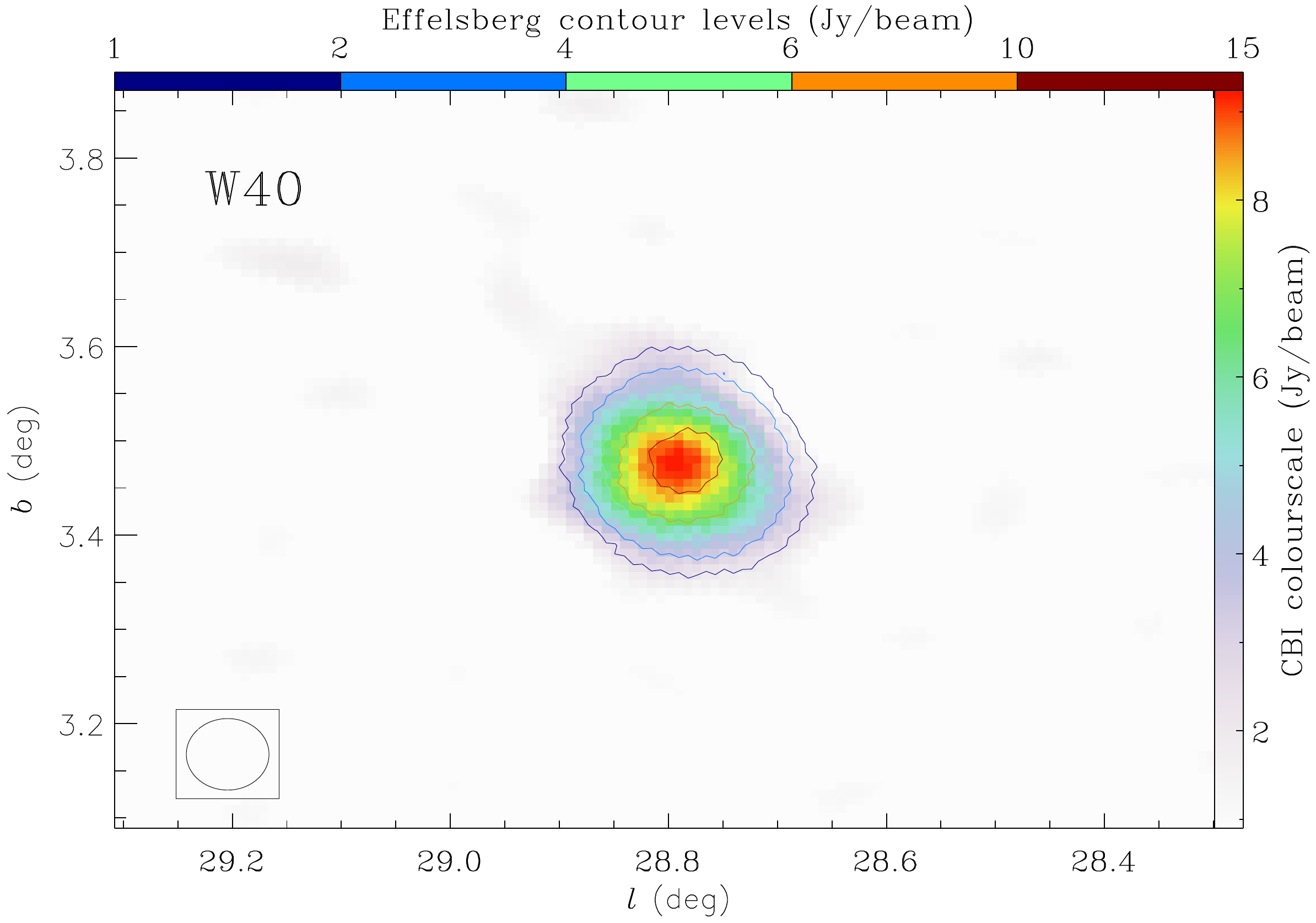}}
%\hfill
\caption[Maps at 31\,GHz of Area-I, Area-II and W40]{In colour-scale are the CBI maps at 31\,GHz of Area-I (left), Area-II (middle) and W40 (right). In the first case over plotted in contours is the simulated-Parkes 5\,GHz map of the region. In the other two cases in contours are the simulated-Effelsberg 2.7\,GHz maps of each region. Both the colour map and the contours are scaled to highlight the fainter sources. The colourbar on the right and top of each map show the brightness levels in the CBI and in the archival maps respectively.  The maps are in units of Jy/beam.}
\label{allareas}
\end{figure*}

\subsection{Ancillary Data}
To study the morphology of the two regions and to understand the emission mechanisms that are in operation, the CBI maps were compared with maps at different frequencies. The ancillary data used for this survey were acquired through the NASA's Skyview\footnote{Skyview: http://skyview.gsfc.nasa.gov/cgi-bin/query.pl, \citet{mcglynn1998}.} and the Max Planck Institute for Radioastronomy survey sampler\footnote{MPIfR sampler: http://www3.mpifr-bonn.mpg.de/survey.html} web services.  Table\,\ref{ancillary} lists the frequencies, angular resolution, units and references for the data. 

\begin{table*}
\begin{center}

\begin{tabular}{c c c c c}
    \hline
    \hline
    Frequency (GHz) & Telescope/Survey & Angular resolution (arcmin) & Units & Reference for the data   \\
    \hline
     0.408 & Haslam & $\sim$\,51 & mK(R-J) & \citet{haslam1982}\\   
    1.4 & NVSS & $\sim$\,0.75 & Jy/Beam &  \citet{condon1998}\\
     2.72 & Effelsberg & $\sim$\,4.3 & mK(R-J) & \citet{furst1990} \\
     4.85 & GB6 & $\sim$\,3.5 & Jy/Beam & \citet{condon1991,condon1993,condon1994}\\
     5.0 & Parkes & $\sim$\,4.1 & mK(R-J) & \citet{haynes1978} \\
     31 & CBI  & $\sim$\,4.5 & Jy/beam & This work \\
    23  & {\it WMAP} & 55.8 & mK(CMB) & \citet{bennett2013}\\
    33  & {\it WMAP} & 40.8 & mK(CMB) & \citet{bennett2013}\\
    41 & {\it WMAP} & 31.8 & mK(CMB) & \citet{bennett2013}\\
    61 & {\it WMAP} & 21 & mK(CMB) & \citet{bennett2013}\\
    94 & {\it WMAP} & 13.8 & mK(CMB) & \citet{bennett2013}\\
    28.4 & {\it Planck} LFI & 33 & mK(CMB) & \citet{planck2013b}\\
    44.1 & {\it Planck} LFI & 24 & mK(CMB) & \citet{planck2013b}\\
    70.4 & {\it Planck} LFI & 14 & mK(CMB) & \citet{planck2013b}\\
    100 & {\it Planck} HFI& 9.5 & mK(CMB) & \citet{planck2013b}\\
    143 & {\it Planck} HFI& 7.1 & mK(CMB) & \citet{planck2013b}\\
    217 & {\it Planck} HFI & 5 & mK(CMB) & \citet{planck2013b}\\
    353 & {\it Planck} HFI & 5 & mK(CMB) & \citet{planck2013b}\\
    545 & {\it Planck} HFI & 5 &  MJy/Sr & \citet{planck2013b}\\
    857 & {\it Planck} HFI & 5 & MJy/Sr & \citet{planck2013b}\\
    2997 (100 ${\rm \mu m}$) & {\it IRAS} & $\sim$\,2 & MJy/Sr & \citet{miville2005} \\
    4995 (60 ${\rm \mu m}$) & {\it IRAS} & $\sim$\,2 & MJy/Sr &  \citet{miville2005}\\
    12875 (25 ${\rm \mu m}$) & {\it IRAS} & $\sim$\,2 & MJy/Sr &  \citet{miville2005}\\
    25750 (12 ${\rm \mu m}$) & {\it IRAS} & $\sim$\,2 & MJy/Sr &  \citet{miville2005}\\
    \hline
    
    \end{tabular}
\end{center}

\caption{Summary of the data used in the study.}
\label{ancillary}
\end{table*}

Archival maps of the two NGC regions at 5\,GHz (after convolving them to the appropriate resolution) are shown as contours in Fig.\,\ref{allareas}. The contours seem to trace the emission of the CBI maps very well. The offset of the brightest sources on the two maps is $\sim$\,60 arcsec and consistent with pointing errors. The diffuse emission is also traced by the contours. This confirms that the data reduction and alignment procedure that was followed performed as expected. It also shows that most of the emission that has been detected by the CBI strongly correlates with the emission at lower frequencies. 

A similar picture was observed when we compared the CBI maps of NGC\,6334 and NGC\,6357 to data at microwave/IR frequencies. The main discrepancy between the CBI and the lower frequency maps is the diffuse emission east of NGC\,6334 detected by the CBI designated as NGC\,6334E (see Fig.\,\ref{allareadiv}). The diffuse emission at 31\,GHz is higher, and has a peak flux at $\sim$\,1.4\,Jy/beam, compared to the emission detected by Parkes which has a peak only at $\sim$\,0.8\,Jy/beam. Further investigation of the NGC\,6334E region showed that the emission detected at 31\,GHz exhibits a strong correlation with the emission in FIR frequencies (for more details see Section\,\ref{6334eanal}).

We also detect a strong correlation between the CBI map and the Effelsberg map at $\sim$\,2.7\,GHz of W51 (see Fig.\,\ref{allareas}). There is some low level diffuse emission, both to the east and north of the cloud, that is visible only to the CBI, due to the different sensitivity of the telescopes. When we compared the W51 map at 31\,GHz to microwave/IR maps we notice that in all cases the bright diffuse emission centred around (\textit{l},{\textit{b})\,=\,(49$\fdg$1,$-$0$\fdg$7), hereafter W51C (see Fig.\,\ref{allareadiv}), disappears. The rest of the emission seen both by the CBI and Effelsberg is traced well by the higher frequency maps. The low level diffuse emission traced only by the CBI is not present at FIR frequencies ($\sim$\,3000\,GHz).

Finally we compared the CBI map of W40 against the 353, 545 and 857\,GHz {\it Planck} maps and the Parkes 5\,GHz map (see Fig.\,\ref{allareas}). We confirm that most of the emission in the area is confined within the $\sim$\,7$'$ region also detected by the CBI. In {\it Planck} maps though the emission is originating from 2 compact like sources around the peak position at 31\,GHz. In addition to that in the {\it Planck} HFI maps we detect diffuse emission, probably originating from cold dust ($T\lesssim15K$), in 2 shell-like regions, surrounding the central W40 area extending up to $\sim$\,1$^{\circ}$ away from the centre. The compact sources and shell-like diffuse emission detected in the HFI {\it Planck} maps does not seem to be present in the {\it IRAS} maps.

\subsection{Measurements and Data fitting}

The incomplete u-v coverage of an interferometer can potentially lead to flux loss at scales greater than the synthesised beam ($>$\,6\,arcmin for the CBI). To allow a reliable comparison to the CBI data, simulated versions of the Parkes 5\,GHz and Effelsberg 2.7\,GHz maps of the four sources of emission were created at the CBI's resolution and u-v coverage. In this way we ensured that any loss of flux density in the CBI maps would also be applied to the ancillary data. To do that the lower frequency maps were first deconvolved with the Richardson-Lucy algorithm. The Richardson-Lucy deconvolution is a Bayesian based iterative procedure for recovering images that were blurred by a known Point Spread Function \citep{richardson1972}. Only a small amount of deconvolution is required since the angular resolution of the CBI, Parkes 5\,GHz and Effelsberg 2.7\,GHz maps is comparable. By comparing a number of deconvolved maps of a given region of the sky we conclude that the process we used is accurate to $\lesssim$1\%. Finally the deconvolved ancillary maps were converted to CBI visibilities using the MOCKCBI software package (for references see Section\,\ref{dataredcal}).

All Area-I, Area-II and W40 measurements were made using aperture photometry. The size and position of each aperture was carefully chosen so that the contamination from nearby sources would be minimised. Additionally bright sources were masked out. Finally the sky contribution was removed. This was done by calculating the median flux density/pixel value within an annulus of an area and inner radius matching the area and radius of the aperture. That value was multiplied by the number of pixels contained in the aperture and was subtracted from the measurement. 

The non-constant spectral distribution of our sources affects the wavelength response of the telescopes' systems. The effect is significant (up to $\sim$\,10\,\%) mainly in the {\it IRAS}, {\it WMAP} and {\it Planck} data due to the large bandwidths of their frequency bands. The method of adjusting the measured flux density to account for that is called colour correcting (CC). To do that we use the publicly available {\it Planck} code and the {\it IRAS} and {\it WMAP} CC tables. 

Using standard propagation of errors we calculate the level of confidence in our measurements. The sources of uncertainties we have taken into account were the CBI calibration errors (see Section\,\ref{dataredcal}), deconvolution accuracy ($\sim$\,1\,\%), uncertainties on the predicted flux at 31\,GHz ($\sim$\,5\,\%) caused by our confidence level on the assumed spectral index, contamination from the CMB fluctuations (found to be negligible for most cases) and nominal errors quoted in the archival data ($\sim$\,10\,\%). Due to the aforementioned sources of uncertainty, on all flux density measurements of NGC\,6357, NGC\,6334, W51 and W40 where CBI data were used we assign a minimum uncertainty of 4.5, 7, 10 and 5 per cent respectively. All measurements containing Parkes, Effelsberg, Haslam and GB6 data extrapolated to 31\,GHz were considered to be accurate to 11.5\,\%. Finally to all flux density measurements containing {\it IRAS}, {\it WMAP} and {\it Planck} data we assign a conservative minimum 3 per cent uncertainty.  Finally the overall uncertainty includes a varying ambiguity in background noise measurements, based on an estimate of the dispersion of the distribution in the background annulus, which was added in quadrature to the uncertainties. 

 To estimate a possible excess emission at $\sim$\,31\,GHz we extrapolated the flux from the lower frequency maps using a power-law fit of the form 
  \begin{equation}
 S=S_{31}\left(\nu _{\rm GHz}/31 \right) ^{\alpha}~, 
 \end{equation}
assuming a constant free-free spectral index $\alpha=-0.12$. This should be a good approximation since we expect that the observed radiation at frequencies $\sim$\,1$-$\,60\,GHz in the absence of Ultra Compact HII (UCHII) regions and AME, to originate from optically thin free-free emission. This type of emission in a typical HII region of our Galaxy has a spectral index  $\alpha\simeq -0.12$ and should not vary by more than $\pm0.02$ for a temperature range $\Delta T$\,$\simeq2000$~K \citep{draine2011}.

An independent investigation for the source of emission for the regions was conducted using the {\it WMAP}, {\it Planck}, {\it IRAS}, GB6, Haslam and Effelsberg maps convolved to $\sim$\,1$^{\circ}$ resolution. 

After measuring the flux $S$ of each source, we fit a model of free-free, synchrotron (where appropriate), thermal dust and spinning dust (where appropriate) components:

\begin{equation}
S=S_{\rm ff}+S_{\rm sync}+S_{\rm td}+S_{\rm sd}~.
\end{equation}

The free-free flux density assuming an electron temperature $T_{\rm e}$=8000\,K was fitted as a power-law
\begin{equation}
S_{\rm ff}=A_{\rm ff}\nu^{\alpha}~,
\end{equation}
where $\alpha$ is an approximation of the free-free spectral index calculated in \citet{draine2011}. This expression is in very good agreement with the theoretical values at frequencies between 0.5 and 10000\,GHz taking into account the steepening of the free-free spectral index at frequencies $\nu \gtrsim100$\,GHz.

The synchrotron component used only for W51 was fitted as a power-law with amplitude $A_{\rm sync}$, and a spectral index $\alpha_{\rm sync}$ calculated using the higher resolution CBI and Effelsberg maps
\begin{equation}
S_{\rm sync}=A_{\rm sync}\nu^{\alpha_{sync}}~.
\end{equation}

The thermal dust was fitted as a modified blackbody 
\begin{equation}
S_{\rm td}=2h\frac{\nu^3}{c^2}\frac{1}{e^{h\nu/KT_{\rm d}}-1}\tau_{250}(\nu/1.2_{\rm THz})^{\beta_{\rm d}}\Omega~,
\end{equation}
where $\tau_{250}$ is the optical depth at 250\,${\rm \mu m}$, $T_{\rm d}$ is the dust temperature, and $\beta_{\rm d}$ is the dust emmisivity index.

 Finally the spinning dust flux density $S_{\rm sd}$ was fitted as a reversed lognormal parabola (Eq.10 in \citealt{bonaldi2007} transformed to flux density units)

\begin{equation}
 S_{\rm sd}=A_{\rm sd} 10^{(F_1(\log\nu)^2+F_2\log\nu)} ~,
\end{equation}
where $F_1$ and $F_2$ are 

\begin{equation}
F_1=\frac{m_{60}}{2\log(\nu_{\rm max}/60GHz)}~,
\end{equation}
\begin{equation}
F_2=\frac{-M_{60}\log(\nu_{\rm max})}{\log(\nu_{\rm max}/60GHz)}~,
\end{equation}

$m$$_{60}$ is the angular coefficient at 60\,GHz and $\nu_{\rm max}$ is the spinning dust peak frequency.

\section{Results}\label{rr}

The resolution of the CBI, Parkes and Effelsberg telescopes ($\sim$\,4$\farcm$5) enabled us to discern the internal structure of all three bright sources of emission in Area-I and Area-II. This allowed us to conduct a separate investigation for each of their components. The division for the three sources is illustrated in Fig.\,\ref{allareadiv}; NGC\,6334 was divided into 6 sub-regions, while both NGC\,6357 and W51 were divided into 4 smaller fields. The names of each sub-region are arbitrary. The dashed line denotes that two measurements were taken for that region, one by having masked out the rest of the sub-regions and thus calculating the diffuse emission in the cloud (thereafter CLEAN), and one without doing that. 

None of the sources' components were distinguishable when we convolved the maps at 1$^{\circ}$ resolution. W40 appears to be a compact source even in our higher resolution maps (see Fig.\,\ref{allareas}); It was therefore treated as a single source throughout the study. 

\begin{figure*}
%\hill
\subfigure{\includegraphics[width=5.76cm]{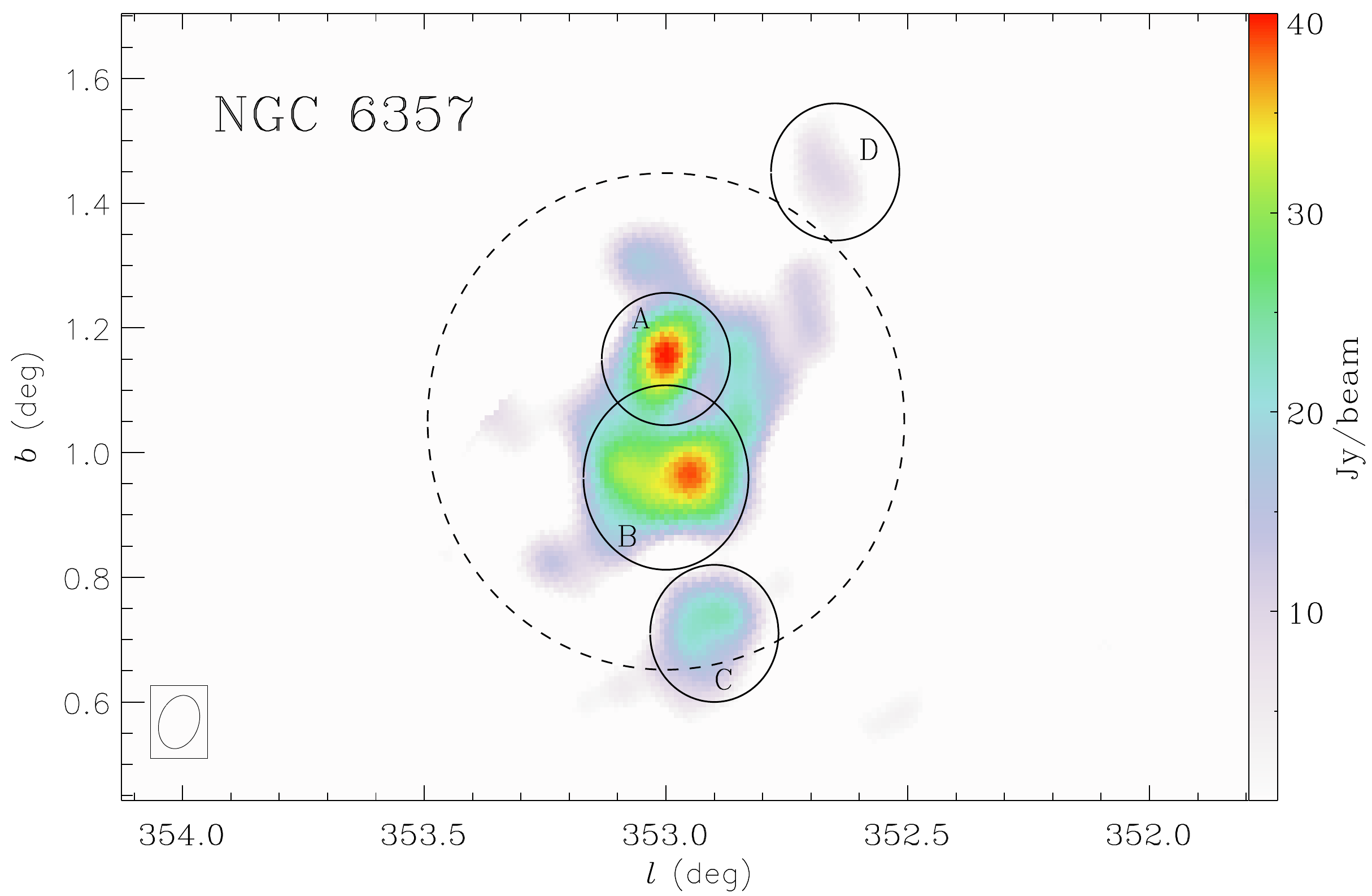}}\hspace{0.5em}
%\hill
\subfigure{\includegraphics[width=5.76cm]{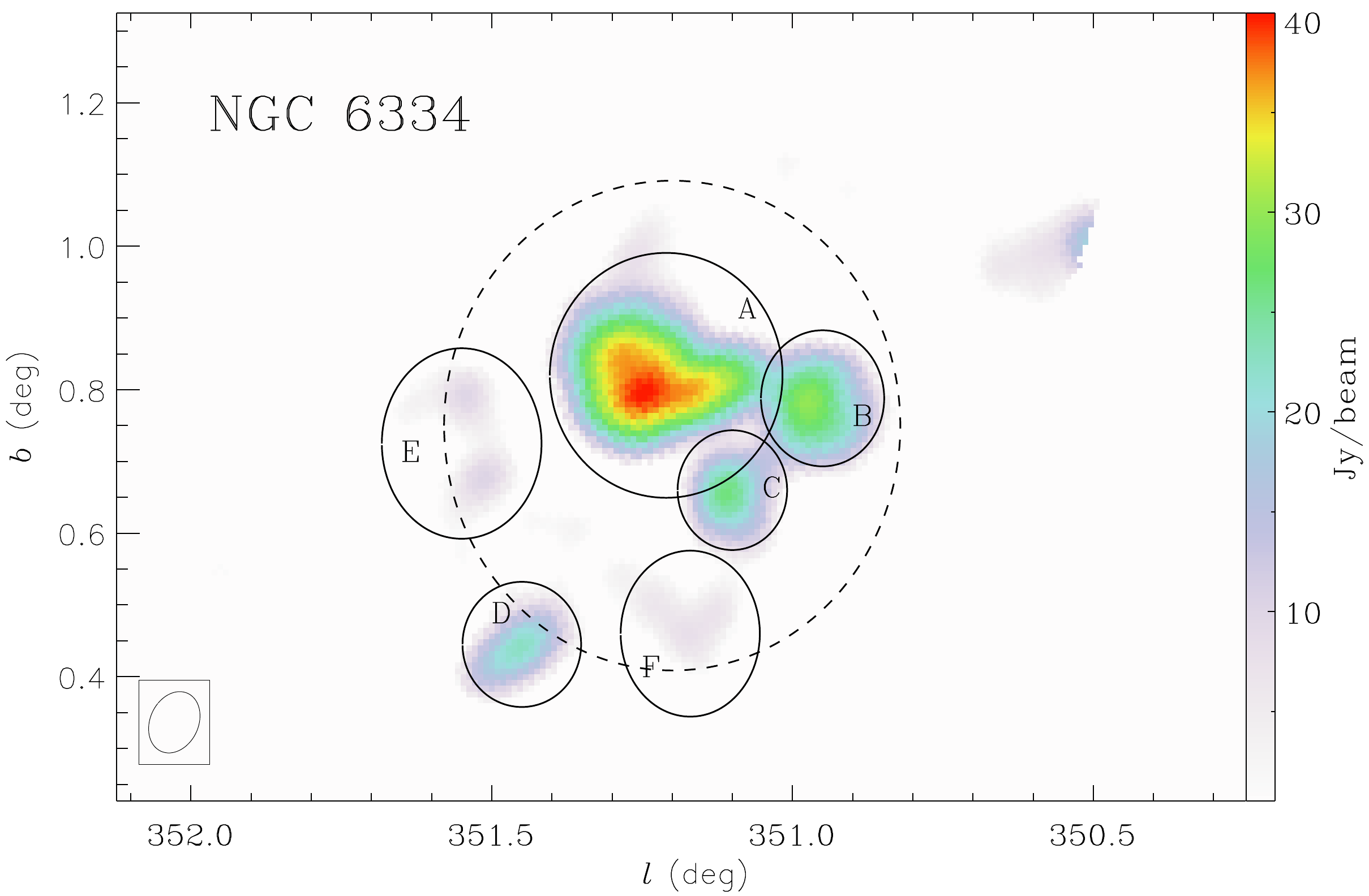}}\hspace{0.5em}
%\hfill
\subfigure{\includegraphics[width=5.76cm]{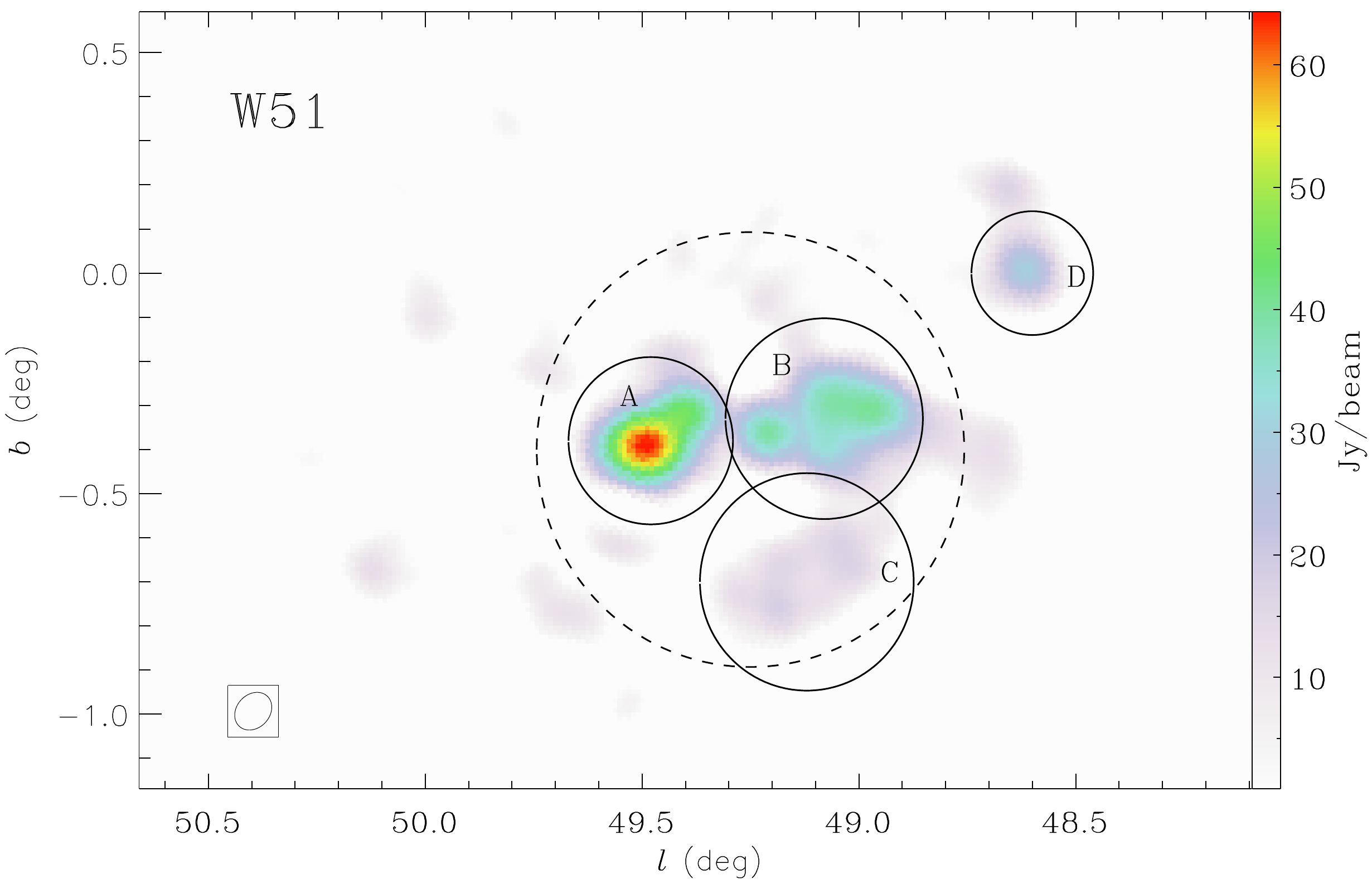}}
%\hfill
\caption[CBI maps at 31\,GHz of NGC\,6357, NGC\,6334 and W51]{CBI maps at 31\,GHz of NGC\,6357 (left), NGC\,6334 (middle) and W51 (right). The colours for the intensity shown on each map were chosen to highlight the faint sources. The peak value in each map is 39.2\,Jy/beam, 30.2\,Jy/beam and 64.9\,Jy/beam. The uniform-weighted beam for each region is $5\farcm4\times 4\farcm2$, $5\farcm4\times 4\farcm2$ and $5\farcm9\times 4\farcm8$, respectively. Over-plotted (continuous and dash lines) are the regions in which by applying photometry we calculated the integrated flux density. The dashed lines denote that two measurements were taken for those regions, one by having masked out the rest of the sub-regions, and one without doing that.}
\label{allareadiv}
\end{figure*}

\subsection{Analysis at 4$\farcm$5 resolution}

\subsection*{NGC\,6357}
The main formations visible in the part of the sky covered by Area-I are NGC\,6357 and NGC\,6334 (Fig.\,\ref{allareas}). \citet{russeil2010} concludes that the extinction and the morphology of the 1.2~mm cold dust emission indicates that both the NGC\,6357 and NGC\,6334 formations are connected by a filamentary structure, therefore belong to a single complex. \citet{neckel1978} and \citet{russeil2012} conclude that both regions are located in the inner edge of the Sagittarius Carina arm at a distance of 1.75 kpc. NGC\,6357 is an extended ($\sim$\,20 pc to 30 pc) region that includes several distinct HII regions in different stages of evolution. At optical wavelengths NGC\,6357 exhibits several bubbles and shell-like regions. At radio wavelengths it is dominated by the two components G353.2+0.9 and G353.1+0.6 (fig.\,1 in \citealt{gvaramadze2011}). Two open clusters, Pismis 24 and AH03 J1725-34.4, are associated with these components respectively \citep{massi1997,gvaramadze2011}. Finally a large number of OB-type stars ($\sim$\,1500) located mainly close to the centre of NGC\,6357 are responsible for the excitation of the ionized nebula \citep{russeil2012}. 

The \citet{russeil2012} study found that there is more 12\,${\rm \mu m}$ emission normalised to the emission at 60\,${\rm \mu m}$ and 100\,${\rm \mu m}$ in NGC\,6357 than in NGC\,6334. The normalised 25\,${\rm \mu m}$ is at a similar level in both clouds. \citet{desert1990} and \citet{galliano2003} found that polycyclic aromatic hydrocarbons (PAHs) dominate the spectrum at wavelengths $ \lambda$\,$\lesssim12$\,${\rm \mu m}$ while very small grains (VSGs) at wavelengths $ \lambda$\,$\sim$\,25\,${\rm \mu m}$. VSG radius ranges between $ 1 \AA \lesssim \alpha_{\rm VSG} \lesssim 10 \AA$, while PAH radius is $\alpha_{\rm PAH} \lesssim 1 \AA$ \citep{galliano2003,galliano2005}. The fact that there are more PAHs in NGC\,6357 than in NGC\,6334 suggests that the former is at a more evolved stage than the latter \citep{russeil2012}. It also suggests that NGC\,6357 may be emitting substantial spinning dust since \citet{haimoud2009} showed that is more likely for the smallest of the VSGs, with a radius $ \alpha \sim$\,0.5$\AA$ (i.e the PAHs), to produce electric dipole emission at the observed by many experiments peak frequency of $\sim$\,30\,GHz.

The fitted and predicted values that were calculated using the $\sim$\,4$\farcm$5 resolution CBI and  Parkes maps, are listed in Table\,\ref{aphrfinalr1}. The data show that in all sub-regions of NGC\,6357 the predicted values for the optically thin free-free emission at 31\,GHz agree with the measured values at the 2$\sigma$ level. The emission in the diffuse component of the cloud (NGC\,6357 CLEAN) is also consistent with optically thin free-free at the 2$\sigma$ level. It is also likely that the overall absolute calibration for the Parkes map could be overestimated as most of the predicted values are slightly higher than the ones measured in the CBI map. Using only the free-free donimated sub-regions we find that the predicted by Parkes flux is on average higher than the CBI by $5\%$. This has been taken into account by assigning a conservative limit of 10\% uncertainties on the Parkes data. We also investigate if AME is associated with cold dust ($T\sim20K$) emitting thermal radiation. The emission from such low temperature objects should peak at FIR frequencies. We therefore convert the limits on excess emission in the NGC\,6357 sub-regions at 31\,GHz to a dust emissivity, relative to the {\it IRAS} 100-${\rm \mu m}$ map. From CMB data at high Galactic latitudes, and at frequencies $\sim$\,30\,GHz, the AME dust emissivity at $\sim$\,30\,GHz has been measured to be $\sim$\,10~$\mu$K\,(MJy sr$^{-1}$)$^{-1}$ with variations around that value of a factor of $\sim$\,2 \citep{Davies2006}. The average value for the emissivity in the sub-regions corresponds to 2.9\,$\pm\,$5.4\,$\mu$K\,(MJy sr$^{-1}$)$^{-1}$ or an upper limit in dust emissivity at the 2$\sigma$ level of $<$~13.7~$\mu$K\,(MJy sr$^{-1}$)$^{-1}$. The upper limit here is therefore consistent with the AME emissivity calculated from high Galactic latitudes, although probably lower due to temperature variations.

\begin{table*}
\footnotesize
 \begin{center}
    \begin{tabular}{| l| p{1.6cm}| p{1.6cm}| p{2.0cm}| p{2.4cm}|p{3.0cm}|}
    \hline
    \hline

     Region  &  Measured $S$$^{5GHz}$~[Jy] & Measured $S$$^{31GHz}$~[Jy]  &  Predicted $S$$^{31GHz}$ [Jy] & Excess $S$$^{31GHz}$~[Jy] &  {\scriptsize Excess~100-${\rm \mu m}$ emissivity} [$\mu$K(MJy sr$^{-1}$)$^{-1}$]   \\
\hline
NGC\,6357 &331\,$\pm$\,38 & 266\,$\pm$\,16 & 266\,$\pm$\,31 & $<$75 & $<$11 \\
\hline
NGC\,6357A &110\,$\pm$\,13 & 80.8\,$\pm$\,4.9 & 88.3\,$\pm$\,10.2 & $<$16.4 & $<$8.8 \\
\hline
NGC\,6357B &152\,$\pm$\,18 & 122\,$\pm$\,7 & 122\,$\pm$\,14 & $<$35 & $<$10 \\
\hline
NGC\,6357C &23.3\,$\pm$\,2.7 & 20.2\,$\pm$\,1.2 & 18.8\,$\pm$\,2.2 & $<$7.0 & $<$27.5 \\
\hline
NGC\,6357D &5.0\,$\pm$\,0.6 & 3.7\,$\pm$\,0.2 & 4.0\,$\pm$\,0.5 & $<$0.8 & $<$9.1 \\
\hline
NGC\,6357 CLEAN &35.1\,$\pm$\,4.0 & 24.3\,$\pm$\,1.5 & 28.2\,$\pm$\,3.2 & $<$3.6 & $<$5.3 \\
\hline
\hline
NGC\,6334 &214\,$\pm$\,25 & 157\,$\pm$\,13 & 172\,$\pm$\,20 & $<$31 & $<$4 \\
\hline
NGC\,6334A &132\,$\pm$\,15 & 97.8\,$\pm$\,8.3 & 106\,$\pm$\,12 & $<$20.8 & $<$4.5 \\
\hline
NGC\,6334B &44.6\,$\pm$\,5.1 & 30.5\,$\pm$\,2.6 & 35.8\,$\pm$\,4.1 & $<$4.1 & $<$5.6 \\
\hline
NGC\,6334C &14.8\,$\pm$\,1.7 & 10.0\,$\pm$\,0.9 & 11.9\,$\pm$\,1.4 & $<$1.2 & $<$8.3 \\
\hline
{\bf NGC\,6334D} &15.6\,$\pm$\,1.8 & 7.2\,$\pm$\,0.6 & 12.5\,$\pm$\,1.4 & $-$5.4\,$\pm$\,1.4 (3.7$\sigma$) & ---- \\
\hline
{\bf NGC\,6334E} &1.6\,$\pm$\,0.2 & 4.0\,$\pm$\,0.3 & 1.3\,$\pm$\,0.2 & 2.7\,$\pm$\,0.4 (6.4$\sigma$) & 9.9\,$\pm$\,1.6 (6.4$\sigma$) \\
\hline
{\bf NGC\,6334F} &2.3\,$\pm$\,0.3 & 1.1\,$\pm$\,0.1 & 1.8\,$\pm$\,0.2 & $-$0.7\,$\pm$\,0.2 (3.3$\sigma$) & ---- \\
\hline
NGC\,6334 CLEAN &5.0\,$\pm$\,0.6 & 3.5\,$\pm$\,0.3 & 4.0\,$\pm$\,0.5 & $<$0.6 & $<$4.3 \\
\hline
   \end{tabular}
   \end{center}
  \caption[Integrated flux densities for the sources in the CBI and simulated Area-I maps, derived using aperture photometry.]{Integrated flux densities for the sources in the CBI and simulated Area-I maps, derived using aperture photometry at 4$\farcm$5 resolution. Errors are quoted at 1$\sigma$ level while the upper limits are given at the 2$\sigma$ level. Regions where the measured flux deviates significantly ($>$3$\sigma$) from the predicted free-free values have boldface names. The notation CLEAN stands for diffuse emission in the cloud.}
\label{aphrfinalr1}
\end{table*}

\subsection*{NGC\,6334}

NGC\,6334 is a bright group of nebulae extending more than $0.\!^{\circ}$5 along the Galactic plane \citep{mcbreen1979}.  At a resolution of $\sim$\,4$\farcm$5 the nebula is resolved in 3 smaller regions; G351.4+0.7 (hereafter NGC\,6334A), G351.0+0.7 (hereafter NGC6334B) and G351.2+0.5 (hereafter NGC6334C). NGC6334A is one of the most active OB star formation region in the Galaxy \citep{loughran1986}. Across the nebula there are a number of extended far-infrared and millimetre sources and radio HII regions that are associated with CO hot spots, some of them having OH and H$_{2}$O masers. Each of the regions is at a different evolutionary stage. Additionally in contrast with NGC\,6357, NGC\,6334 is ionized by a small number of OB stars that are spread around the whole nebula \citep{tapia1996,russeil2012}.

In the results of the investigation that was carried out using the CBI and simulated Parkes maps of NGC\,6334 (see Table\,\ref{aphrfinalr1}), unlike to NGC\,6357, a complex picture emerges. Although for the brightest parts of the nebula (NGC\,6334A, NGC\,6334B, NGC\,6334C) and its diffuse component (NGC\,6334 CLEAN) the measured flux density agrees with predicted values for an optically thin free-free dominated region, for the sub-regions NGC\,6334D, NGC\,6334E and NGC\,6334F this is not the case. Relative to a single power-law with slope $\alpha$\,=\,$-$0.12 used to extrapolate the flux from 5 to 31\,GHz we detected excess emission in NGC\,6334E at $\sim$\,6.4$\sigma$ level and emission deficiency in NGC\,6334D and NGC\,6334F at $\sim$\,3.7$\sigma$ and $\sim$\,3.3$\sigma$ respectively. The excess emission at 31\,GHz for the rest of the NGC\,6334 sub-regions were converted to a dust emissivity, relative to the {\it IRAS} 100-${\rm \mu m}$ map. This corresponds to an emissivity of $-$1.8\,$\pm$\,2.3\,$\mu$K\,(MJy sr$^{-1}$)$^{-1}$ or a 2$\sigma$ upper limit of $<$\,2.8~$\mu$K\,(MJy sr$^{-1}$)$^{-1}$. For more discussion about the dust emissivity in the clouds see Section\,\ref{ameem}.

\subsection*{NGC\,6334D and NGC\,6334F}
For the three NGC\,6334 sub-regions we have found that optically thin free-free is not the only emission mechanism that is active, thus a separate investigation is required. We first focus on the two sub-regions where the emission deficit has been detected. Using the measured flux densities at 5 and 31\,GHz we calculate the value for the spectral index of NGC\,6334D and NGC\,6334F and find it to be  $\alpha= -0.42\pm0.08$ and $\alpha= -0.39\pm0.11$ respectively.

NGC\,6334F is the SNR G351.2+0.1. Its spectral index at 1\,GHz has been measured to be  $-0.4$ \citep{caswell1975,whiteoak1996}, consistent with our findings. The lack of any significant spectral breaks denotes that a power-law is a good model for the SED of the source for frequencies between 1 and 30\,GHz. The investigation for the source of emission in NGC\,6334D focussed on finding SNR candidates by using the D. A. Green SNR catalogue\footnote{Green catalogue: http://www.mrao.cam.ac.uk/surveys/snrs.}\citep{green2014}, and finding Galactic and extragalactic sources that would have a steep spectral index by looking at the Set of Identifications, Measurements and Bibliography for Astronomical Data SIMBAD\footnote{SIMBAD: http://simbad.u-strasbg.fr/simbad/.} \citep{wenger2000} and at the Nasa Extragalactic Database NED\footnote{NED: http://ned.ipac.caltech.edu/forms/nearposn.htm. The NASA/IPAC Extragalactic Database (NED) is operated by the Jet Propulsion Laboratory, California Institute of Technology, under contract with the National Aeronautics and Space Administration.}. We did not find any synchrotron emission sources in that region of the Galaxy that could account for the emission that has been observed. Looking in the higher resolution 1.4\,GHz NVSS map of the region we found two point like sources with peak values at 2.1\,Jy/beam and 2.0\,Jy/beam. NGC\,6334D is also visible on all {\it IRAS} and { \it Herchel} maps of the region.

The calibration errors alone could not have altered our findings of NGC\,6334D by more than 10\%. Primary beam errors even as far as 2 times the FWHM of the telescope could not have affected the flux density value of the source by more than 20\% either (see left panel of fig.\,5 in \citealt{taylor2011}). The emission deficit in the region can be accounted for if NGC\,6334D is a hotter HII region ($T$=15000\,K), and if with the aforementioned sources of errors working together are conspiring to force the spectral index to become steeper.  Finally in order to cover all possibilities we speculate that the source is likely to be an unknown SNR, or a distant galaxy with a steep spectral index that happens to lie close to the Galactic plane.

\subsection*{NGC\,6334E}\label{6334eanal}

As mentioned in Section\,\ref{dataim}, NGC\,6334E is a 10 by 10 arcmin$^2$ area east of the main NGC\,6334 formation (see Fig.\,\ref{ngc6334e}). At this point we can not be certain if it is part of the cloud or it happened to lay on the line-of-sight either closer or further away from it. Most of the emission in the sub-region is originating from two compact components, which are connected  with a filament. More low level emission is being emitted from the two sources extending further to their east and south-east. Although there is faint to no emission in the sub-region at 5\,GHz (see Table.\ref{aphrfinalr1}), NGC\,6334E is quite bright in the IR. When we compared the CBI map of the sub-region against maps at higher frequency we find that there is a strong correlation with the IRAS 25\,${\rm \mu m}$ map (shown as contours in Fig.\,\ref{ngc6334e}), and the Spitzer 24\,${\rm \mu m}$ map indicating that the excess emission we have detected might be spinning dust. There is an offset though of $\sim$\,1.5' between the two maps. This offset has been observed in previous CBI studies as well. We could have performed a separate investigation for each of the two sources, but due to their small size (smaller than the beam of the CBI), their proximity to each other and to the rest of NGC\,6334 we chose to consider them as one source.

\begin{figure} 
\centering
\includegraphics[width=3.5in]{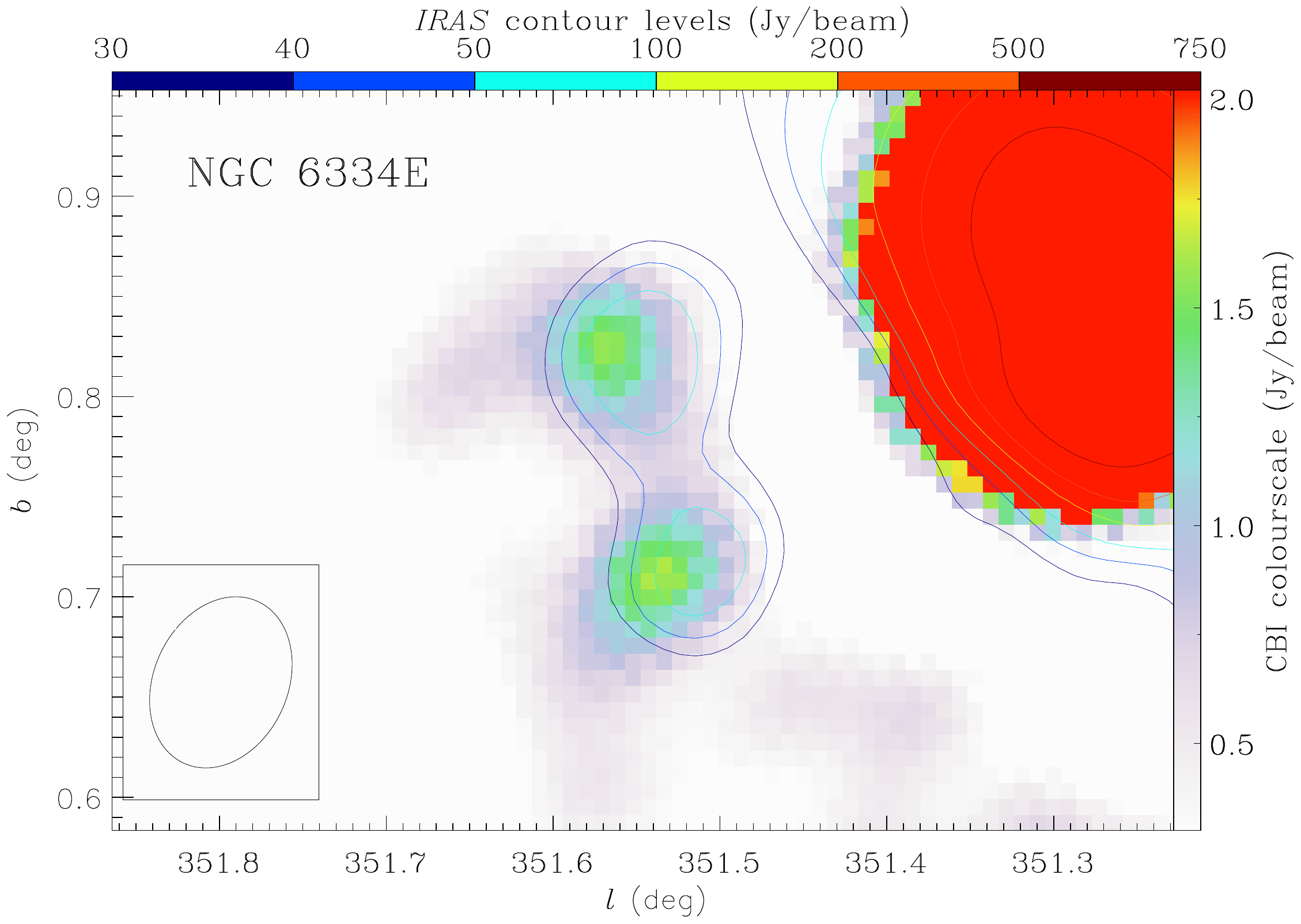} 
\caption[CBI map of NGC\,6334E.]{CBI map of NGC\,6334E at 31\,GHz and at $\sim$\,4$\farcm$5 resolution. Overplotted in contours is the IRAS 25\,${\rm \mu m}$ map convolved at the same resolution and sampled in the CBI u-v coverage.}
\label{ngc6334e}
\end{figure}

The excess emission detected at the 6.4$\sigma$ level in NGC\,6334E clearly shows that the dominant emission mechanism in the region at frequencies between 5 and 31\,GHz is not optically thin free-free. Using the measured flux densities at 5 and 31\,GHz we calculate the spectral index of the region to have a value of $\alpha=0.50\pm0.06$. The excess emission accounts for $\sim$\,70\,\% of the total emission of the sub-region at 31\,GHz. The additional emission that has been observed could originate from the AME mechanism or from free-free emission coming from UCHII regions.

HII regions with an EM  $> 10^7$ cm pc$^{-6}$ (UCHII regions) will produce free-free emission that is optically thick at frequencies below $\sim$\,15\,GHz \citep{kurtz1994}. In the optically thick regime the flux in the region will increase as $\sim$\,$\nu^2$, while at higher frequencies it will fall off like $\sim$\,$\nu^{-0.1}$. This could mean that although faint at frequencies lower than 15\,GHz, these sources could exhibit flux densities of up to $\sim$\,10\,Jy at $\sim$\,30\,GHz. It is therefore possible that the AME (or a portion of it) that was detected in NGC\,6334E could be produced by UCHII regions. There are several methods for calculating the contribution of UCHII in the overall flux of a region. One way involves using high resolution radio data at frequencies below 15\,GHz, to identify the point sources in the region. After measuring their EM and assuming an angular size for the sources we extrapolate the flux density at higher frequencies. We looked at the NRAO VLA survey NVSS\footnote{NVSS~URL: http://www.cv.nrao.edu/nvss/} at 1.4\,GHz \citep{condon1998} and the Co-Ordinated Radio 'N' Infrared Survey for High-mass star formation CORNISH\footnote{CORNISH~URL: http://cornish.leeds.ac.uk/public/index.php} at 5\,GHz\,\citep{hoare2012,purcell2013} to verify if any UCHII are present in the NGC\,6334E region, find their optical depth and estimate their contribution to the overall flux that was measured in the CBI final maps at 31\,GHz. The optical depth of the brightest source in NGC\,6334E at 1.4\,GHz, assuming a $T_{\rm e}=8000$K, is $\tau << 1 \simeq 0.002$, i.e we found no significant UCHIIs in the region. Unfortunately the CORNISH survey does not cover this part of the sky. An alternative method, proposed by \citet{dickinson2013}, is to look for UCHII candidates by looking at the colour-colour relation of \citet{Wood1989}, who found that UCHII regions tend to have IRAS colour ratios of ${\rm log}_{10}(S_{60}/S_{12}) \geq 1.30$ and ${\rm log}_{10}(S_{25}/S_{12})\geq 0.57$. Then by using the empirical relation between the ratio of the flux at 100\,${\rm \mu m}$ and 15\,GHz \citep{kurtz1994} and a free-free spectral index we extrapolate the flux at 31\,GHz \citep{dickinson2013}. We looked for sources that matched the aforementioned criteria within 10$'$ of 351.693+0.579. We found no such sources in the region surrounding NGC\,6334E. Both tests leave AME as the most plausible candidate.  If NGC\,6334E is part of the NGC\,6334 cloud it means that AME can be found localised on a part of the cloud where the conditions are right, even if the rest of the structure is free-free dominated.

Such special conditions for AME to be active can be met for example in photodissociation regions (or photon-dominated regions, PDRs, \citealt{casassus2008,tibbs2012b}). A PDR is a region of interstellar gas that is sufficiently cold to remain neutral, but due to its low column density far-UV photos can penetrate it hence heat it and influence its chemistry. According to current star formation theories PDRs usually form around HII regions, in shell-like structures, separating the ionised and the molecular gas. PAHs are mostly confined in the surrounding PDRs, while the warmer dust is mainly located at the centre of these shell-like regions. Comparing the {\it Spitzer} maps of NGC\,6334E at 8\,${\rm \mu m}$ (tracing the PAH population) and 24\,${\rm \mu m}$ (tracing the VSGs) we found that no such morphological conditions are present in the sub-region. We then compared the PAH/VSG ratio of NGC\,6334E to the rest of the cloud. To do that we normalised the {\it Spitzer}\,8\,${\rm \mu m}$ emission of each area to the corresponding {\it IRAS} 25\,${\rm \mu m}$ emission. We use the {\it IRAS}\,25\,${\rm \mu m}$ map as a tracer for the VSGs because there are no {\it Spitzer} data for the NGC\,6334 cloud other than for NGC\,6334E. The 8/24 flux (and i.e PAH/VSG) ratio for NGC\,6334 and NGC\,6334E was found to be $\sim$\,0.25 and $\sim$\,0.95 respectively. This much higher PAH/VSG ratio in NGC\,6334E compared to the rest of the cloud indicates that the whole sub-region might be a PDR of NGC\,6334.

Finally we use the {\it Planck} maps at 147, 353, 454 and 857\,GHz and the {\it IRAS} 100\,${\rm \mu m}$ map (after sampling them to the u-v coverage of the CBI) to fit for the thermal dust optical depth of NGC\,6334E at 1.2\,THz ($\tau_{250}$). We then normalise the excess emission measured in the CBI map with the optical depth at 1.2\,THz to calculate the AME emissivity of the sub-region at 31\,GHz relative to $\tau_{250}$. We find it to be equal to (1.0\,$\pm$\,0.2)$\times 10^4$\,Jy. For further discussion about AME emissivities in the clouds see Section\,\ref{ameem}.

\subsection*{W51}

In the second data-set (Area-II) one important formation is visible, the W51 complex. W51 is a massive molecular complex located at the tangential point of the Sagittarius arm of the Galaxy at a distance of  $\sim$\,5.5\,kpc \citep{sato2010}. W51, as seen in the radio band images, can be divided into three main components. W51A and W51B are two star forming regions that form what is called the W51 giant molecular cloud (W51~GMC), which spreads across an area of 1\,deg$^2$. W51C is a diffuse and extended component that is attached to the south-eastern boundary of W51B and expands to the east. W51C was identified as a supernova remnant (SNR) and is estimated to be around 30$\times 10^3$ years old \citep{koo1995}. Finally the existence of two 1720 MHz OH masers and the detection of about 100 solar masses of atomic gas at a velocity shifted between 20 and 120 kms$^{-1}$  with respect to its ambient medium \citep{green1997,koo1997} suggest that there is interaction between the two regions, W51B and W51C.

The results for the investigation that was carried out using the CBI and Effelsberg maps are shown in Table\,\ref{aphrfinalr2}. The predicted optically thin free-free and measured integrated flux densities of W51A, W51B, W51D and W51 CLEAN are in agreement at the 2$\sigma$ level. The predicted optically thin free-free overall flux for the region is higher though than the one measured in the CBI maps by $\sim$\,15\,\% (1$\sigma$). This could be either due to calibration errors (see Section\,\ref{dataredcal}) or due to the additional synchrotron emission at low radio frequencies that was recorded in W51C at the 6.5$\sigma$ significance level.

\begin{table*}
\footnotesize
    \begin{tabular}{| l| p{1.6cm}| p{1.6cm}| p{2.0cm}| p{2.2cm}|p{3.0cm}|}
    \hline
    \hline
    Region  &  Measured $S$$^{2.7GHz}$~[Jy] & Measured $S$$^{31GHz}$~[Jy]  &  Predicted $S$$^{31GHz}$ [Jy] & Excess $S$$^{31GHz}$~[Jy] &  {\scriptsize Excess~100-${\rm \mu m}$ emissivity} [$\mu$K(MJy sr$^{-1}$)$^{-1}$]   \\
    
\hline
W51 &354\,$\pm$\,41 & 231\,$\pm$\,23 & 264\,$\pm$\,30 & $<$\,37 & $<$\,5 \\
\hline
W51A &150\,$\pm$\,17 & 117\,$\pm$\,12 & 112\,$\pm$\,13 & $<$\,37 & $<$\,9 \\
\hline
W51B &95.5\,$\pm$\,11.0 & 63.8\,$\pm$\,6.4 & 71.3\,$\pm$\,8.2 & $<$\,11.6 & $<$\,5.5 \\
\hline
{\bf W51C} &49.0\,$\pm$\,5.6 & 11.7\,$\pm$\,1.2 & 36.6\,$\pm$\,4.2 & $-$24.9\,$\pm$\,3.8 (6.5$\sigma$) & ---- \\
\hline
W51D &9.5\,$\pm$\,1.1 & 7.5\,$\pm$\,0.8 & 7.1\,$\pm$\,0.8 & $<$\,2.5 & $<$\,8.0 \\
\hline
W51 CLEAN &40.5\,$\pm$\,4.7 & 25.0\,$\pm$\,2.5 & 30.2\,$\pm$\,3.5 & $<$\,2.6 & $<$\,2.6 \\
\hline
\hline
W40 &31.5 \,$\pm$\,3.4 & 28.1\,$\pm$\,1.7 & 23.5 \,$\pm$\,2.5 & $<$\,10.6 &  $<$\,8.2 \\
\hline

    \end{tabular}

  \caption[Integrated flux densities for the sources on the CBI and Effelsberg Area-II and W40 maps, derived using aperture photometry.]{Integrated flux densities for the sources on the CBI and Effelsberg Area-II and W40 maps, derived using aperture photometry at 4$\farcm$5 resolution. Errors are quoted at 1$\sigma$ level while the upper limits are given at the 2$\sigma$ level. W51C deviates significantly ($>$3$\sigma$) from the predicted optically thin free-free and is shown in boldface. The notation CLEAN stands for diffuse emission in the cloud.}
\label{aphrfinalr2}
\end{table*}

We can once again convert the limits on excess emission in the sub-regions of the W51 (besides W51C) at 31\,GHz to a dust emissivity, relative to the {\it IRAS} 100\,${\rm \mu m}$ map (Table\,\ref{aphrfinalr2}). The values that were calculated are considerably lower though than the values that were found in diffuse regions at high Galactic latitudes.
The mean dust emissivity of the sub-regions is $-$1.5\,$\pm\,$3.3\,$\mu$K\,(MJy sr$^{-1}$)$^{-1}$ which corresponds to a 2$\sigma$ limit of $<5.1$\,$\mu$K\,(MJy sr$^{-1}$)$^{-1}$.

\subsection*{W51C}
The quoted spectral index of W51C at 1\,GHz is $\alpha$=$-$0.3 \citep{shaver1970,velusamy1974,subrahmanyan1995}, confirming that the dominant emission mechanism in the region is synchrotron radiation. W51C is a type S SNR with an age of $\sim$\,3\,$\times 10^4$\,yr \citep{hanabata2013,park2013}. Using CBI and Effelsberg data we calculate a spectral index of $\alpha$=$-$0.58$\pm$0.06. This value indicates spectral ageing of the more energetic electrons resulting in a steeper index at about a few GHz.

Assuming now that all the emission in W51C at 31\,GHz is due to free-free we extrapolate to 2.7\,GHz to find S$_{{ \rm free-free}}^{2.7{\rm GHz}}$=$15.7 \pm 1.6$\,Jy. We then subtract that value from the measured flux of W51C at 2.7\,GHz to find S$_{{\rm sync}}^{2.7{\rm GHz}}$=$34.0 \pm 5.9$\,Jy. We therefore conclude that at least $\sim$\,70\,\% of the flux of W51C and $\simeq$\,10\% of W51 at 2.7\,GHz is due to synchrotron radiation.

\subsection*{W40}

W40 is an HII complex $\sim$\,8$'$ in diameter. A number of energy sources reside inside this cloud, with the most prominent being three young and large OB stars giving rise to ionising UV radiation \citep{zeilik1978}. \citet{rodney2008} found a total of 20 radio sources, half of which presented fluctuations in their flux over time. Their results were recently confirmed by VLA observations in radio frequencies \citep{rondriguez2010}. The same authors speculate on the nature of these sources suggesting that they may be combination of ultra compact HII regions, young stellar objects (YSOs) and shocked interstellar gas. Finally, a thick dust cloud heavily reduces the visibility of W40 to such an extent that we cannot view a significant portion of W40 in the optical and IR bands \citep{reyle2002,rondriguez2010}. The distance to W40 has not been well determined yet. Estimates give values between 0.3 and 0.9 kpc \citep{radhakrishnan1972}. The mass of the cloud has been estimated to be around 10$^4 M_{\odot}$ \citep{zhu2006}. 

Assuming a free-free spectral index $\alpha$=$-0.12$ we find that the measured fluxes at 2.7\,GHz and 31\,GHz agree at the 2$\sigma$ level (see Table\,\ref{aphrfinalr2}). We therefore conclude that at $\sim$\,68\,\% and $\sim$\,95\,\% confidence limits that the AME in the region is S$^{AME}_{1\sigma} <$7.6\,Jy and S$^{AME}_{2\sigma}<$10.6\,Jy, respectively. Once again the excess emission was converted to dust emissivity relative to the {\it IRAS} maps 100\,${\rm \mu m}$, returning a value of 3.7\,$\pm$\,2.4\,$\mu$K(MJy sr$^{-1}$)$^{-1}$.

\subsection{Analysis at 1$^{\circ}$ resolution}

The Spectral Energy Distributions (SEDs) at 1$^{\circ}$ scale for NGC\,6357, NGC\,6334, W51 and W40 are shown in Fig.\,\ref{SEDallsource}. Solid black circles show the {\it Planck} data at 100 and 217\,GHz. These points were identified as being consistently higher than the rest due to contamination by strong CO line emission at 115 and 230\,GHz, so they were not considered during the parameter fitting stage \citep{planck2013c}. Table\,\ref{lwmapr12} summarises the results. From the four plots we can infer that there is a small excess emission in all 4 regions at 31\,GHz although too faint to be definitively detected. It might be that AME is present in all 4 cases or that there is an unknown source of systematics in the data. We do not expect though that systematics of that level would significantly bias our results. 

For NGC\,6357 we find that our fitted values for the optical depth at 1.2~THz $\tau_{250}=(23.0\pm3.8)\times 10^4$, the dust temperature $T$$_{\rm d}=21.3 \pm 0.7$\,K and the dust emissivity $\beta_{\rm d}=1.81 \pm 0.07$ are consistent at the 2$\sigma$ level with the results of \citet{planck2014}.  We also find that at 28.4\,GHz the overall flux density of NGC\,6357 is consistent with a free-free emission at the 2$\sigma$ level ($\Delta S^{28.4}$\,=\,$45\pm35$\,Jy). The emissivity of the cloud at 28.4\,GHz relative to $\tau_{250}$ is $(1.5\pm 1.2) \times 10^4$\,Jy, a value also consistent at the 2$\sigma$ level with the results of \citet{planck2014}.
 
 \begin{figure*}
%\hill
\subfigure{\includegraphics[width=8.0cm]{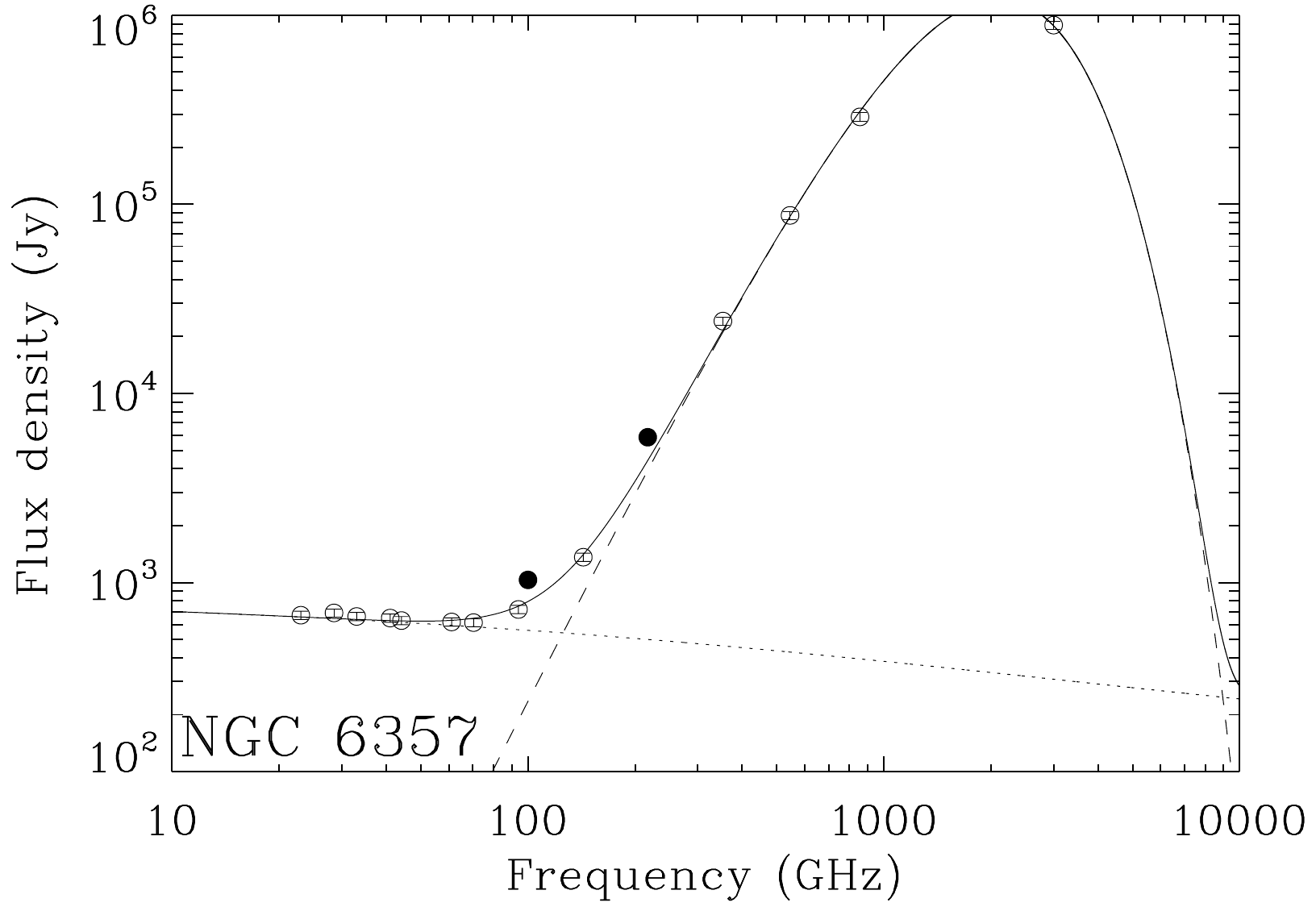}}\hspace{0.5em}
%\hill
\subfigure{\includegraphics[width=8.0cm]{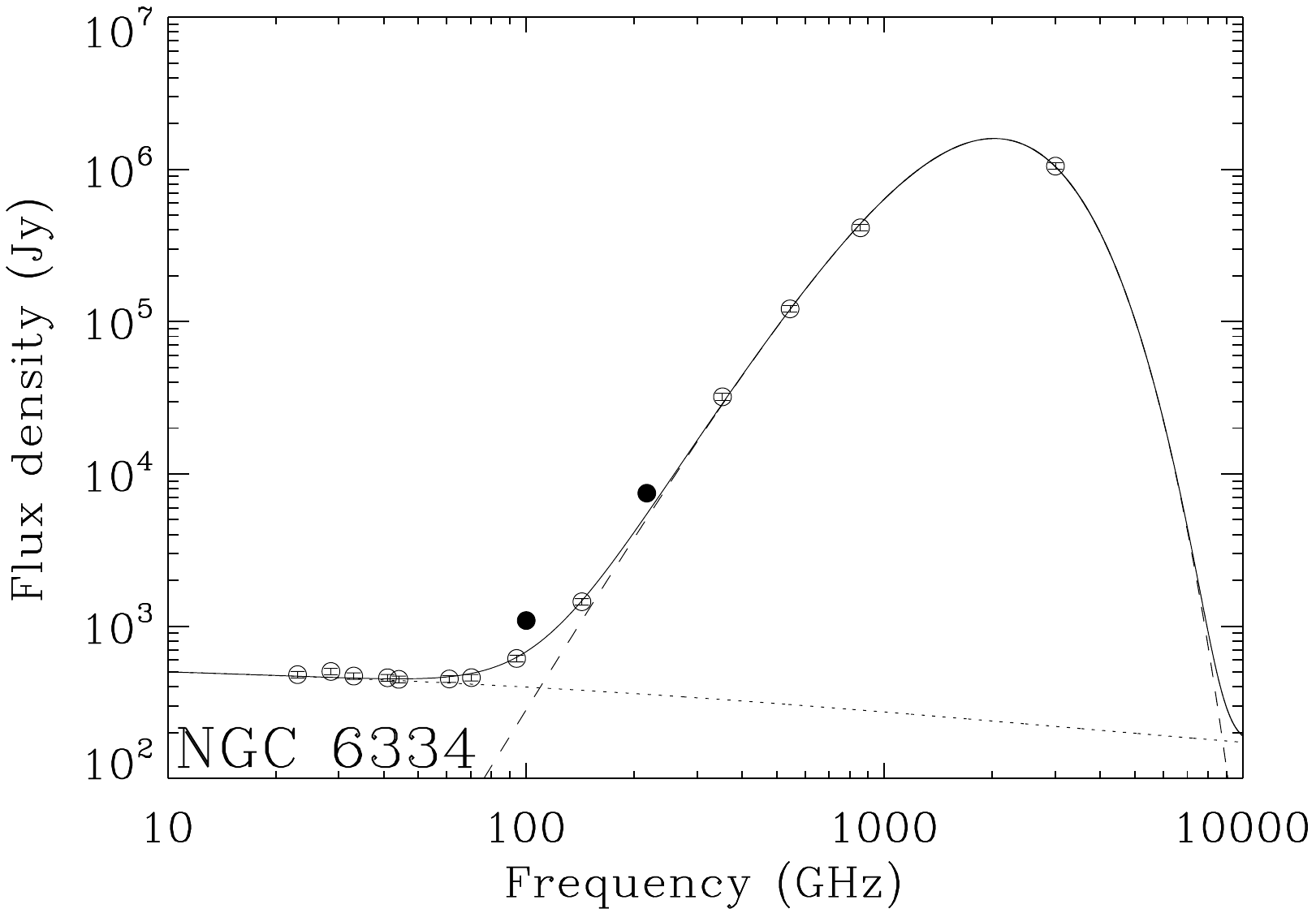}}\hspace{0.5em}
%\hfill
\subfigure{\includegraphics[width=8.0cm]{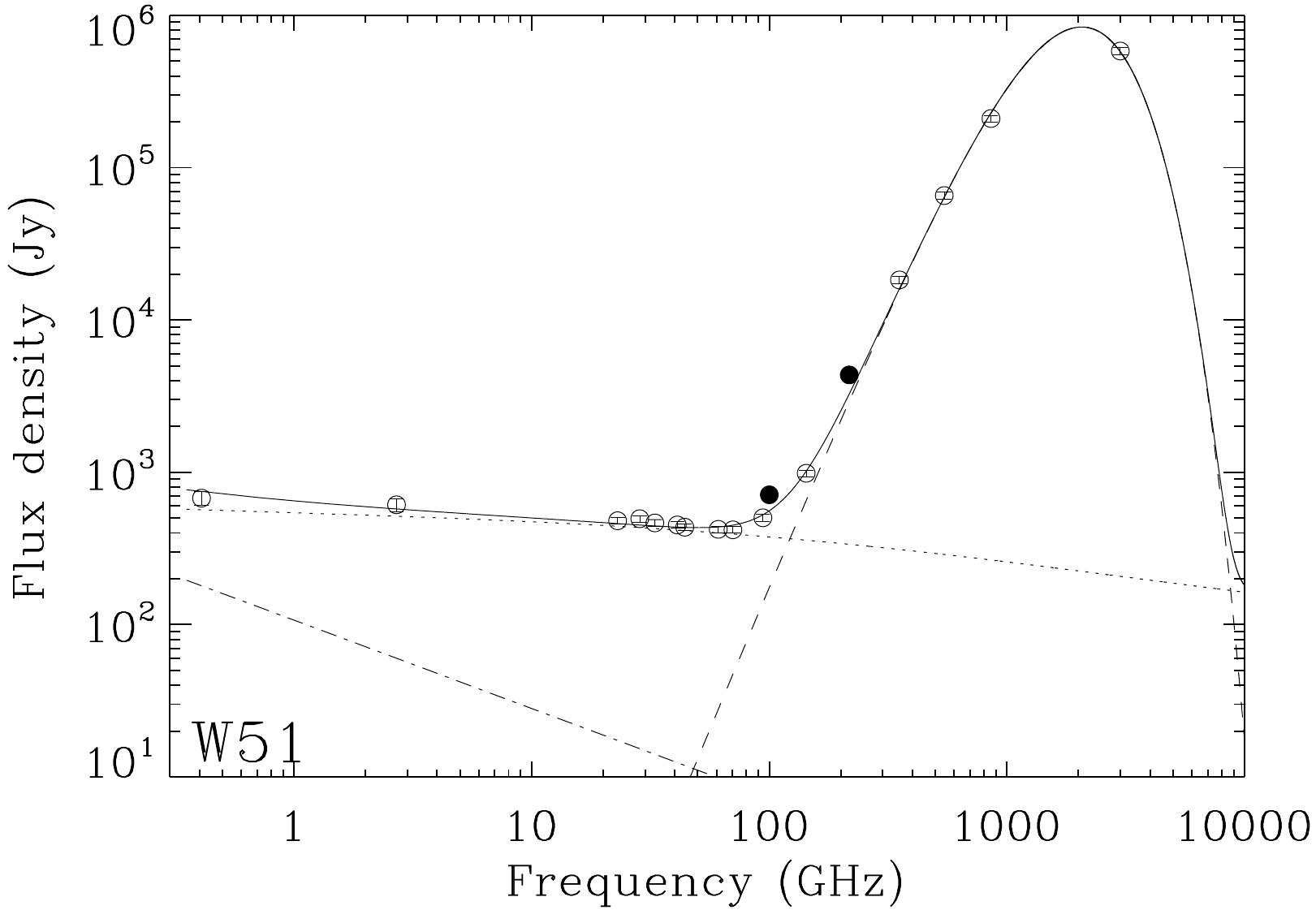}}\hspace{0.5em}
%\hfill
\subfigure{\includegraphics[width=8.0cm]{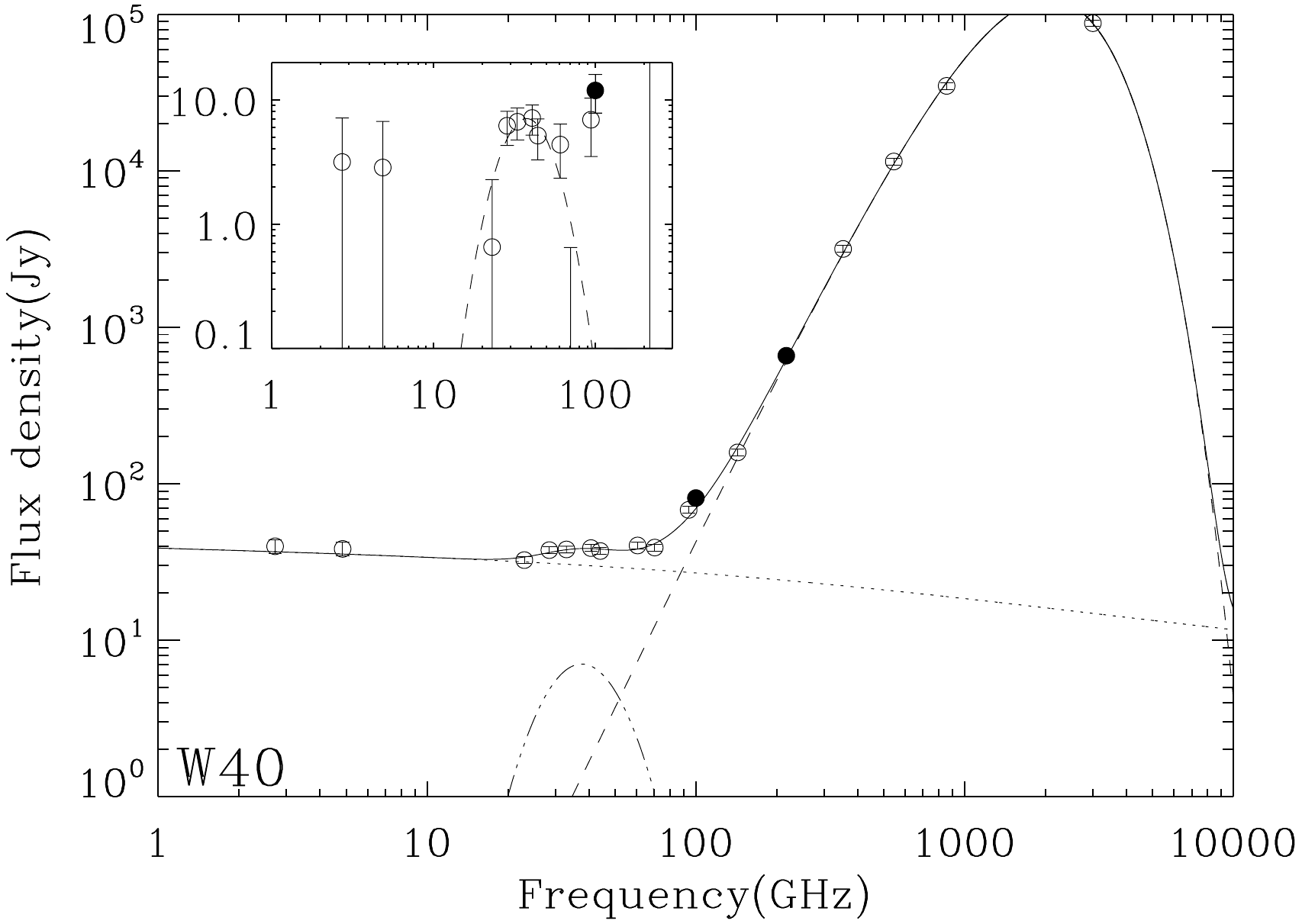}}
%\hfill
\caption[Spectral energy distributions of NGC\,6357, NGC\,6334, W51 and W40 at 1$^{\circ}$ resolution.]{Spectral energy distributions of NGC\,6357 (top-left), NGC\,6334 (top-right), W51 (bottom-left) and W40 (bottom-right) at 1$^{\circ}$ resolution. A conservative limit of 3\% uncertainties was used to fit the data. The uncertainties include calibration errors and the fact that the CMB anisotropies are no longer negligible at these scales and frequencies. In all four cases the spectrum was fitted using a model for free-free (dotted line) and thermal dust (dashed line). Additionally the SED of W51 includes a synchrotron emission model (dot-dash line) and of W40 a model for the spinning dust (treble dot-dashed line). The sub-figure of the W40 spectral energy distribution shows the residual flux density in each data point after subtracting the free-free and synchrotron components of the spectrum. Overplotted (dashed line) is the fitted to the data spinning dust model.}
\label{SEDallsource}
\end{figure*}

\begin{table*}
  \begin{center}
   \begin{tabular}{|c| c| c| c| c| c| c| c| c| c|}  
    \hline
    \hline
    Region & $\tau_{250} \times 10^{4}$ & $T$$_{\rm d} [K]$ & $\beta_{\rm d}$ & $\Delta S^{28.4}$ [Jy] & $\Delta S^{28.4}$/$\tau_{250}\times 10^{-4}$ [Jy] & $\sigma_{\rm AME}^{28.4}$ & $\sigma_{\rm AME}^{cumul}$ & A$_{\rm sd}$ [Jy] & $\nu_{ \rm sd}$ [GHz]\\
    \hline
    NGC\,6334 & 24.1\,$\pm$\,3.8 & 19.9\,$\pm$\,0.5 & 1.93\,$\pm$\,0.06 & 43.2\,$\pm$\,25.2 & 1.4\,$\pm$\,0.8 & 1.7 & -- & -- & --\\
    \hline
    NGC\,6357  & 23.0\,$\pm$\,3.8 & 21.3\,$\pm$\,0.7 & 1.81\,$\pm$\,0.07 & 45.1\,$\pm$\,34.5& 1.5\,$\pm$\,1.2 & 1.3 & -- & -- & --\\
    \hline
    W51     & 11.3\,$\pm$\,2.3 & 20.7\,$\pm$\,0.8 & 1.82\,$\pm$\,0.09  & 60.1\,$\pm$\,25.0 & 4.1\,$\pm$\,1.8 & 2.3 & -- & -- & --\\
    \hline
    W40  & 3.9\,$\pm$\,0.3 & 21.7\,$\pm$\,0.6 & 1.60\,$\pm$\,0.06 & 6.2\,$\pm$\,1.9 & 1.6\,$\pm$\,0.5 & 3.3 & 6.4 & 7.1 $\pm$ 2.8 & 37.6 $\pm$ 4.3\\
    \hline
    \end{tabular}
  \end{center}
  \caption[A list of the fitted parameters to the SEDs of the four regions at 1$^{\circ}$ resolution.]{A list of the fitted parameters to the SEDs of the four regions at 1$^{\circ}$ resolution. Uncertainties are quoted at 1$\sigma$ level. Note that $\sigma_{\rm AME}^{28.4}$ and $\sigma_{\rm AME}^{cumul}$ are the AME detection significance at $28.4$\,GHz and cumulative detection of AME in all frequency channels between 10 and 100\,GHz.}
\label{lwmapr12}
\end{table*}

For NGC\,6334 the results indicate that the cloud is dominated by optically thin free-free in the frequency range of 20-80\,GHz and if additional radiation is emitted from any part of the nebula, is not significant enough (AME significance = 1.3$\sigma$) to make a difference in the overall flux that we have measured. The fitted values for $\tau_{250}$, T$_{\rm d}$ and $\beta_{\rm d}$ are $(24.1\pm3.8)\times 10^4$, $19.9 \pm 0.5$\,K and $1.93 \pm 0.06$ respectively. The limits on excess emission at 28.4\,GHz were once again used to calculate the emissivity relative to $\tau_{250}$ corresponding to a value of $(1.5\pm1.2) \times 10^4$\,Jy.

The SED of W51 is shown in the bottom-left panel of Fig.\,\ref{SEDallsource}. To account for the synchrotron emission that was detected earlier we additionally use the Haslam 408\,MHz map and the Effelsberg 2.7\,GHz and fit a power-law with a fixed spectral index $\alpha=-0.58$ (value calculated using the CBI and Effelsberg maps). The results show that synchrotron accounts for $\sim$\,25\% of the emission of the cloud at 408\,MHz, $\sim$\,10\% at 2.7\,GHz, and is negligible at higher frequencies. These findings are in good agreement with our results on W51C at 4.5$'$ resolution. We also fit for $\tau_{250}$, $T$$_{\rm d}$ and $\beta_{\rm d}$ and find that they are equal to $(11.3 \pm 2.3) \times 10^4$, $20.7\pm0.8$\,K and $1.82\pm 0.09$ respectively. Finally we calculated the emissivity at 28.4\,GHz relative to $\tau_{250}$ and found it to be equal to $(4.1 \pm 1.8) \times 10^4$\,Jy.

The SED of W40 calculated at $\sim$\,1$^{\circ}$ resolution is shown in Fig.\,\ref{SEDallsource}. The fitted values for $\tau_{250}$, $T$$_{\rm d}$ and $\beta_{\rm d}$ are $(3.9\pm0.3)\times 10^4$, $21.7 \pm 0.6$\,K and $1.60 \pm 0.06$ respectively. To better constrain the free-free component in the region, in addition to the {\it WMAP} and {\it Planck} maps, we use the Effelsberg map at 2.7\,GHz and the GB6 map at 4.8\,GHz. We detect a slightly rising spectrum at frequencies between $\sim$\,20 and 40\,GHz where the spinning dust emission mechanism is expected to be active. There is on average an additional $\sim$\,20\,\% more emission than what it was anticipated.  At 28.4\,GHz AME was detected at the 3.3$\sigma$ level, while the cumulative AME detection between 10 and 100\,GHz was at the 6.4$\sigma$ level. These findings are in agreement with \citet{finkbeiner2004b} who noted an excess at $\sim$\,30\,GHz. 

An investigation for UCHII regions within a 30$'$ radius from W40, using the IRAS data and the colour-colour relation of \citealt{Wood1989} (see Section\,\ref{6334eanal}) revealed two UCHII candidates. The estimated upper limit on the combined flux density of the two sources at $\nu=28.4$\,GHz, based on the empirical relation between the flux density of a source at 100\,$\mu$m and 15\,GHz \citep{kurtz1994} and a free-free spectral index $\alpha$\,=\,-$0.12$, is $\sim0.25$\,Jy. This value accounts for $<5$\% of the detected excess emission in the area. Using the NVSS data at 1.4\,GHz to identify point sources around W40 we found one potential UCHII region. The derived optical depth of the source, assuming an electron temperature of $T_e$\,=\,$8000$\,K, is $\tau$\,$\simeq$\,0.01\,$<<$\,1. Unfortunately no CORNISH data of the region are available. Our study has shown that no significant contribution from UCHII regions is expected at frequencies $\sim$\,30\,GHz leaving the spinning dust emission mechanism as the only known valuable alternative.

We therefore fit a spinning dust component to the spectral energy distribution of W40 (shown as a treble dot-dashed line in the SED of W40 in Fig.\,\ref{SEDallsource}, and emphasised in the sub-figure of the same plot) which in turn reduces the value of $\chi^2$ by $\sim$\,15 with only 3 additional parameters. The fitted flux density and peak frequency are $A_{\rm sd}$\,=\,7.1\,$\pm$\,2.8\,Jy ($\sim$\,2.7$\sigma$) and $\nu_{\rm sd}\,=\,$37.6\,$\pm$\,4.3\,GHz. These values are also in good agreement with \citet{finkbeiner2004b}. \citet{planck2014} on the other hand detected an AME significance in the region of only 0.2$\sigma$. The results depend significantly on the two higher uncertainty lower frequency data, not used by the Planck collaboration, since they are the ones constraining the level of free-free in the region. The values constrained by the thermal dust component of the SED on the other hand ($\tau_{250}$, $T$$_{\rm d}$ and $\beta_{\rm d}$), are consistent with the values calculated in \citet{planck2014} at the 1$\sigma$ level. The results are also consistent with our CBI-Efflesberg study of the region at the 1$\sigma$ level. Finally we calculated the AME emissivity at 28.4\,GHz relative to the optical depth at 1.2\,THz, $\tau_{250}$ and found it to be (1.6\,$\pm$\,0.5)\,$\times 10^4$\,Jy.

\subsection{AME emissivities}\label{ameem}

The unweighted mean value of the dust emissivity relative to the {\it IRAS}-100\,${\rm \mu m}$ maps of the 14 sub-regions of NGC\,6334, NGC\,6357 and W51 and of W40 is $0.5\pm4.4$\,$\mu$K(MJy sr$^{-1}$)$^{-1}$. At the 2$\sigma$ confidence level this is equivalent to $<$\,$9.3$\,$\mu$K(MJy sr$^{-1}$)$^{-1}$. This value is somewhat lower than the $10$\,$\mu$K(MJy sr$^{-1}$)$^{-1}$ calculated in cooler regions \citep{Davies2006}. This lower ratio is more likely to arise, if AME is due to the spinning dust emission mechanism, due to the fact that the hotter HII region environments destroy the smaller grains which give a rise in the spectrum at frequencies $\sim$\,30\,GHz \citep{povich2007}. It could also be due to the fact that the emissivity drops by a factor of $\sim$\,10 when we move from 20 to 30~K \citep{tibbs2012}. It is more appropriate to calculate the emissivity relative to other standards, like the optical depth, which is an indicator of the column density relative to hydrogen $n_{\rm H}$ \citep{finkbeiner1999,finkbeiner2004b}. 

Calculating the AME emissivity of the 4 regions at 28.4\,GHz relative to the optical depth at 1.2\,THz, $\tau_{250}$, was found to always lie on the order of $\sim$\,$10^4$\,Jy, and appears to be independent of the AME significance (see Table\,\ref{lwmapr12}). This suggests that the AME is proportional to the dust column density independant of how strong the AME source is, in agreement with the results of \citet{planck2014}. The unweighted and weighted average AME emissivity of the four regions relative to dust column density is $(2.1\pm 1.3) \times 10^4$\,Jy and $(1.9\pm 0.6) \times 10^4$\,Jy respectively. The AME emissivity of NGC\,6334E at 31\,GHz relative to $\tau_{250}$ was also calculated to be equal to $(1.0 \pm 0.2)\times 10^4$. The unweighted and weighted AME emissivities quoted in \citet{planck2014} are  $(4.3 \pm 0.6)\times 10^4$ and $(1.4 \pm 0.1)\times 10^4$ respectively, being in agreement with our results at the 2$\sigma$ level.

\section{Conclusions}\label{concl}

In this paper we present the results of the investigation for the source of emission for the Galactic regions NGC\,6357, NGC\,6334, W51 and W40. CBI observations of the four regions were conducted at $\sim$\,4$\farcm$5 resolution.

We have confirmed both using high resolution ($\sim$\,4$\farcm$5) interferometric data and single dish data convolved at 1$^{\circ}$ resolution that the source of emission at frequencies between 20 and 80\,GHz in NGC\,6357 is consistent with optically thin free-free at the 2$\sigma$ level. For frequencies $\gtrsim$100\,GHz thermal dust takes over in NGC\,6357 as in the rest of the clouds.

The main source of emission in NGC\,6334 at 1$^{\circ}$ resolution and in the frequency range of 20-80\,GHz is optically thin free-free. Using the higher resolution interferometric data though we detected that at 31\,GHz, 3 of the cloud's sub-regions, designated as NGC\,6334D, NGC\,6334E and NGC\,6334F were inconsistent with the extrapolated from lower frequency data optically thin free-free, at the 2$\sigma$ level. NGC\,6334D and NGC\,6334F show a lack of emission relative to the predicted optically thin free-free at 3.3$\sigma$ and 3.7$\sigma$ respectively. NGC\,6334F is a known SNR with a spectral index of $-$0.4 consistent with values from the literature. We could not identify a source (or group of sources) that match the spectral characteristics found in NGC\,6334D. We therefore conclude that the sub-region might be an unknown SNR.

We also detected an excess emission compared to the predicted free-free in NGC\,6334E at 6.4$\sigma$. We ruled out the possibility that the excess emission is due to UCHII regions. We calculated the AME emissivity for the sub-region at 31\,GHz relative to the optical depth at 1.2\,THz and find it to be equal to (1.0$\pm$0.2)\,$\times 10^4$\,Jy. The PAH/VSG ratio in NGC\,6334E compared to the rest of NGC\,6334 suggests that the former might be a PDR of the latter.

The emission detected in all sub-regions of W51 other than W51C at frequencies between 20 and 80\,GHz is due to optically thin free-free. W51C on the other hand is an SNR with a spectral index of $-$0.58. Synchrotron accounts for $\gtrsim$70\% of the measured flux of W51C and $\sim$\,10\,\% of the W51 cloud at 2.7\,GHz. The results are consistent both using the higher resolution interferometric data and the lower resolution single dish data. 

Although optically thin free-free is the dominant emission mechanism in W40 in the frequency range of 20-80\,GHz we detected a spinning dust component at $\sim$\,3.3$\sigma$, that accounts up to 20\% of its total emission at the peak frequency $\nu_{\rm sd}\simeq37$\,GHz. 

Finally we calculated the AME emissivity at 31\,GHz of all sub-regions of the clouds relative to maps at 100\,${\rm \mu m}$. We find an average value of $0.5\pm4.4$\,$\mu$K(MJy sr$^{-1}$)$^{-1}$. This value is lower than the 10\,$\mu$K(MJy sr$^{-1}$)$^{-1}$ measured at high Galactic latitudes \citep{Davies2006}, probably due to temperature variations. We also calculated the AME emissivity of the 4 regions at 28.4\,GHz, and the AME emissivity of NGC\,6334E at 31\,GHz relative to the optical depth at 1.2\,THz. This form of emissivity in all 4 regions and the NGC\,6334E sub-region lies somewhere on the order of $10^4$\,Jy, and is independent of the AME significance, similar to recent findings of \citet{planck2014}.

AME is a significant component of the spectrum at frequencies around 30\,GHz, for a number of sources in our Galaxy. Additionally AME seems to be present mainly in diffuse parts of the Galaxy. NGC\,6334E should be followed-up at higher resolution and at frequencies $\sim$\,30\,GHz. 

\section*{Acknowledgements}

This work was supported by the Strategic Alliance for the Implementation of New Technologies (SAINT -\
 see www.astro.caltech.edu/chajnantor/saint/index.html) and we are most grateful to the SAINT partners for their strong support. We gratefully acknowledge support from the Kavli
  Operating Institute and thank B. Rawn and S. Rawn Jr. The CBI was supported by NSF grants 9802989, 0098734
  and 0206416, and a Royal Society Small Research Grant. We are
  particularly indebted to the engineers who maintained and operated the
  CBI: Crist{\'o}bal Achermann, Jos{\'e} Cort{\'e}s, Crist{\'o}bal Jara, Nolberto Oyarace, Martin Shepherd and Carlos Verdugo. 
  
Constantinos Demetroullas is grateful to Anna Bonaldi for clarifying the equation for fitting spinning dust and to Matias Vidal for his valuable comments and support throughout many stages of this study. Constantinos also acknowledges the support of a STFC quota studentship and a President's Doctoral Scholarship from the University of Manchester and he is grateful to the European Research Council for support through the award of an ERC Starting Independent Researcher Grant (EC FP7 grant number 280127). Clive Dickinson is supported by an STFC Consolidated Grand (no.~ST/L000768/1), an EU Marie Curie Re-Integration Grant and an ERC Starting Grant (No. 307209). Constantinos and Clive would like to thank NVSS, CORNISH, NED, SIMBAD and Green catalogue.

\bibliographystyle{mnras}
\bibliography{cbipaper}

\label{lastpage}

\end{document}